\documentclass[12pt,english,aps,superscriptaddress,preprintnumbers]{revtex4-1}
\usepackage[T1]{fontenc}
\usepackage[latin9]{inputenc}
\usepackage{float}
\usepackage{mathrsfs}
\usepackage{amsmath}
\usepackage{amssymb}
\usepackage{graphicx}
\usepackage{esint}

\makeatletter

\providecommand{\tabularnewline}{\\}

\usepackage{mathrsfs}

\@ifundefined{textcolor}{}{%
 \definecolor{BLACK}{gray}{0}
 \definecolor{WHITE}{gray}{1}
 \definecolor{RED}{rgb}{1,0,0}
 \definecolor{GREEN}{rgb}{0,1,0}
 \definecolor{BLUE}{rgb}{0,0,1}
 \definecolor{CYAN}{cmyk}{1,0,0,0}
 \definecolor{MAGENTA}{cmyk}{0,1,0,0}
 \definecolor{YELLOW}{cmyk}{0,0,1,0}
}

\usepackage{babel}

\makeatother

\usepackage{babel}
\begin{document}
\title{Thermodynamics of Chiral Fermion System in a Uniform Magnetic Field}
\author{Cheng Zhang}
\email{These authors contributed equally to this work.}

\affiliation{Key Laboratory of Particle Physics and Particle Irradiation (MOE),
Institute of Frontier and Interdisciplinary Science, Shandong University,
Qingdao, Shandong 266237, China}
\author{Ren-Hong Fang}
\email{These authors contributed equally to this work.}

\affiliation{Key Laboratory of Particle Physics and Particle Irradiation (MOE),
Institute of Frontier and Interdisciplinary Science, Shandong University,
Qingdao, Shandong 266237, China}
\author{Jian-Hua Gao}
\email{Corresponding authors. houdf@mail.ccnu.edu.cn (D.-F.H.), gaojh@sdu.edu.cn (J.-H.G.)}

\affiliation{Shandong Provincial Key Laboratory of Optical Astronomy and Solar-Terrestrial
Environment, Institute of Space Sciences, Shandong University, Weihai,
Shandong, 264209, China}
\author{De-Fu Hou }
\email{Corresponding authors. houdf@mail.ccnu.edu.cn (D.-F.H.), gaojh@sdu.edu.cn (J.-H.G.)}

\affiliation{Institute of Particle Physics and Key Laboratory of Quark and Lepton
Physics (MOS), Central China Normal University, Wuhan 430079, China}
\begin{abstract}
We construct the grand partition function of the system of chiral
fermions in a uniform magnetic field from Landau levels, through which
all thermodynamic quantities can be obtained. Taking use of Abel-Plana
formula, these thermodynamic quantities can be expanded as series
with respect to a dimensionless variable $b=2eB/T^{2}$. We find that
the series expansions of energy density, pressure, magnetization intensity
and magnetic susceptibility contain a singular term with $\ln b^{2}$,
while particle number density, entropy density and heat capacity are
power series of $b^{2}$. The asymptotic behaviors of these thermodynamic
quantities in extreme conditions are also discussed.
\end{abstract}
\maketitle

\section{Introduction}

The properties of matter under the electromagnetic field have been
studied extensively these years in many fields of physics. It is well
known that strong electric field can lead to pair production of fermions
in QED vacuum, which is called Schwinger mechanism \citep{Schwinger:1951,Kim:2002,Brezin:1970}.
Recently the effect of magnetic field on Schwinger mechanism is studied
through the approaches of equal-time Wigner function and AdS/CFT correspondence
\citep{Zhang:2017egb,Sheng:2018jwf,Zhu:2019igg}. In astronomy, compact
stars, such as white dwarfs, neutron stars, and quark stars, often
rotate very rapidly, which can produce magnetic field as strong as
$10^{12}\sim10^{15}$ Gauss \citep{Felipe:2007vb,Reisenegger:2003pj}.
This strong magnetic field may have great impact on the state of compact
stars \citep{Itokazu:2018lij,Reisenegger:2013faa}. In high energy
physics, such as peripheral high energy heavy ion collisions, there
also produce strong magnetic field in the collision region \citep{Skokov:2009qp,Bzdak:2011yy,Voronyuk:2011jd,Deng:2012pc},
and may induce the currents of charged particles along the direction
of the magnetic field, which is called chiral magnetic effect \citep{Kharzeev:2007jp,Fukushima:2008xe,Gao:2012ix,Feng:2018tpb,Liang:2020sgr}.
In condensed matter physics, the strong magnetic field can reduce
chiral condensate, which is called magnetic catalysis \citep{Gusynin:1994xp,Fukushima:2012xw,Bali:2013esa,Miransky:2015ava,Mao:2018wqo,Ballon-Bayona:2020xtf}.
Meson condensation under background of the magnetic field together
with the electric field or rotation is also discussed in \citep{Cao:2015cka,Fang:2016uds,Chen:2019tcp,Wang:2018gmj}.
The magnetic field also has an important influence on the thermodynamics
and transport properties of the system of massive fermions \citep{Cangemi:1996tp,Huang:2009ue}.
In relativistic hydrodynamics, the Wigner function approach is often
used to study the hydrodynamics of fermion system in general electromagnetic
field \citep{Gao:2012ix,Chen:2012ca,Hidaka:2016yjf,Gao:2018wmr,Yang:2020mtz}.
The chiral kinetic theory in the electromagnetic field is also studied
recently \citep{Gorbar:2017cwv,Lin:2019fqo}. 

In this article, we study the influence of magnetic field on the thermodynamics
of the chiral fermion system, where we ignore the interaction among
the fermions. Since the equations of motion for left-handed and right-handed
fermions decouple, we will only consider the case of right-handed
fermions in this article, and all results can be generalised to the
left-handed case directly. In the previous work \citep{Dong:2020zci}
by some of us, the electric current of the right-handed fermion system
along the magnetic field, which is explained as chiral magnetic effect,
has been obtained through the ensemble average of normal ordering
of corresponding operator. In this article we will use the method
of grand partition function instead, from which we can obtain all
knowledge of the thermodynamic system. Through solving the stationary
Schroedinger equation of a single right-handed fermion in a uniform
magnetic field, we can obtain a series of Landau levels, from which
we can construct the grand partition function. According to the standard
procedure in quantum statistical mechanics, all thermodynamic quantities
can be obtained by the grand partition function. In fact, for the
thermodynamic system of massive fermions, the proper-time method is
the most popular method to calculate the grand partition function
\citep{Dittrich:1979ux,Gies:1998vt,Ozaki:2015yja}. However, for the
thermodynamics of chiral fermion system in this article, we will take
use of the Abel-Plana formula to calculate the grand partition function.
In the expression of the grand partition function there is a summation
over all Landau levels, which is difficult to be dealt with analytically.
Fortunately, there is a Abel-Plana formula which can transform the
discrete summation into integrations. Then we can express the grand
partition function as a two-dimensional integration, which can be
expanded as a series with respect to a dimensionless variable $b=2eB/T^{2}$,
with $e$ the electric charge of the right-handed fermion, $B$ the
magnetic field and $T$ the temperature of the system. In the series,
besides the terms with $b^{2n}$, there is an additional singular
term $b^{2}\ln b^{2}$, which indicates that the grand partition function
is not analytic at $b=0$, leading to the non-analyticity of some
thermodynamic quantities at $b=0$. We investigate the asymptotic
behaviors of these thermodynamic quantities in the extreme conditions,
such as weak/strong magnetic field limits and high/low temperature
limits. Our study of the effect of the magnetic field on the thermodynamics
of the chiral fermion system may have important theoretical meaning
for the research on the quark gluon plasma state which can be produced
in high energy heavy ion collisions.

The rest of this article is organised as follows. In Sec. \ref{sec:Landau levels},
the Landau levels of a single right-handed fermion in a uniform magnetic
field is briefly listed. In Sec. \ref{sec:partition function}, we
construct the grand partition function from Landau levels and express
all thermodynamic quantities by the grand partition function. In sec.
\ref{sec:Expansion}, all thermodynamic quantities are expanded as
series with respect to a dimensionless variable $b$. In Sec. \ref{sec:-extreme},
we study the asymptotic behaviors of these thermodynamic quantities
in extreme conditions. In Sec. \ref{sec:tensor}, all elements of
the energy-momentum tensor are calculated. This article is summarized
in Sec. \ref{sec:Summary}.

Throughout this article we adopt natural units where $\hbar=c=k_{B}=1$.
The convention for the metric tensor is $g^{\mu\nu}=\mathrm{diag}\,(+1,-1,-1,-1)$.
We use the Heaviside-Lorentz convention for electromagnetism and the
chiral representation for gamma matrixes where $\gamma^{5}=\mathrm{diag\,(-1,-1,+1,+1)}$,
which is the same as Peskin and Schroeder \citep{Peskin:1995}. 

\section{Landau levels for a single right-handed fermion in a uniform magnetic
field}

\label{sec:Landau levels}The Lagrangian of a chiral (massless) fermion
field $\psi$ under the background of a uniform magnetic field $\boldsymbol{B}=B\boldsymbol{e}_{z}$
is
\begin{equation}
\mathcal{L}=\bar{\psi}i\gamma\cdot D\psi,\label{eq:5.1a}
\end{equation}
where $D^{\mu}=\partial^{\mu}+ieA^{\mu}$, with $e$ the electric
charge of the fermion and $A^{\mu}$ the gauge potential chose as
$A^{\mu}=(0,0,Bx,0)$. In this article we set $eB>0$ for simplicity.
The results of all thermodynamic quantities in this article can be
extended to the range $eB<0$.

In the chiral representation of gamma matrixes, we can write $\psi=(\psi_{L},\psi_{R})^{T}$,
where the two-component spinors $\psi_{L}$ and $\psi_{R}$ are called
left-handed and right-handed fermion fields. Euler-Lagrange equation
of the Lagrangian in Eq. (\ref{eq:5.1a}) gives
\begin{equation}
i\frac{\partial}{\partial t}\psi_{L}=-i\boldsymbol{\sigma}\cdot\boldsymbol{D}\psi_{L},\label{eq:5.1b}
\end{equation}
\begin{equation}
i\frac{\partial}{\partial t}\psi_{R}=i\boldsymbol{\sigma}\cdot\boldsymbol{D}\psi_{R},\label{eq:5.1c}
\end{equation}
where $\boldsymbol{\sigma}=(\sigma^{1},\sigma^{2},\sigma^{3})$ are
Pauli matrixes and $\boldsymbol{D}=(-\partial_{x},-\partial_{y}+ieBx,-\partial_{z})$.
Since the equations of motion for $\psi_{L}$ and $\psi_{R}$ decouple,
we only discuss right-handed fermion field in this article. All results
can be directly generalised to the left-handed case. 

The stationary Schroedinger equation $i\boldsymbol{\sigma}\cdot\boldsymbol{D}\psi_{R}=E\psi_{R}$
gives a series of Landau levels and eigenfunctions as follows \citep{Dong:2020zci},
\begin{eqnarray}
n=0, & E=k_{z}, & \psi_{R0}(k_{y},k_{z};\boldsymbol{x})=\left(\begin{array}{c}
\varphi_{0}(\xi)\\
0
\end{array}\right)\frac{1}{L}e^{i(yk_{y}+zk_{z})},\nonumber \\
n>0, & \ E=\lambda E_{n}(k_{z}),\  & \psi_{Rn\lambda}(k_{y},k_{z};\boldsymbol{x})=c_{n\lambda}\left(\begin{array}{c}
\varphi_{n}(\xi)\\
iF_{n\lambda}\varphi_{n-1}(\xi)
\end{array}\right)\frac{1}{L}e^{i(yk_{y}+zk_{z})},\label{eq:5.1e}
\end{eqnarray}
where $\lambda=\pm1$, $\xi=\sqrt{eB}x-k_{y}/\sqrt{eB}$, $E_{n}(k_{z})=\sqrt{2neB+k_{z}^{2}}$,
$F_{n\lambda}=\left[k_{z}-\lambda E_{n}(k_{z})\right]/\sqrt{2neB}$,
$|c_{n\lambda}|^{2}=1/(1+F_{n\lambda}^{2})$, and $\varphi_{n}(\xi)$
is the $n$-th harmonic oscillator function along $x$-axis whose
center is $x=k_{y}/eB$. We have assumed that the eigenfunctions are
set up in a box with sides of lengths $L$, i.e., $0<x,y,z<L$, and
satisfy periodic boundary conditions in the $y$-axis and $z$-axis,
i.e., $k_{y}=2\pi n_{y}/L$, $k_{z}=2\pi n_{z}/L$, $(n_{y},n_{z}=-\infty,\cdots,\infty)$.
The condition that the center of the oscillation along $x$-axis is
inside the box leads to $0<n_{y}<eBL^{2}/(2\pi)$. Since the energy
level $E_{n}(k_{z})$ is independent of $k_{y}$, the degeneracy of
each Landau level is $eBL^{2}/(2\pi)$. 

\section{Grand partition function and thermodynamic quantities}

\label{sec:partition function}We consider a system of right-handed
fermions in a uniform magnetic field $\boldsymbol{B}=B\boldsymbol{e}_{z}$,
which is in equilibrium with a reservoir with temperature $T$ and
chemical potential $\mu_{R}$ . The interaction among the fermions
in this system is ignored for simplicity. From the Landau levels for
a single right-handed fermion in Sec. \ref{sec:Landau levels}, we
can construct the grand partition function $\ln\Xi$ of this system
as follows,
\begin{eqnarray}
\ln\Xi & = & \sum_{k_{y},k_{z}}\bigg[\theta(k_{z})\ln(1+e^{a-\beta k_{z}})+\theta(-k_{z})\ln(1+e^{-a+\beta k_{z}})\bigg]\nonumber \\
 &  & +\sum_{n=1}^{\infty}\sum_{k_{y},k_{z}}\bigg[\ln(1+e^{a-\beta\sqrt{2neB+k_{z}^{2}}})+\ln(1+e^{-a-\beta\sqrt{2neB+k_{z}^{2}}})\bigg],\label{eq:5.1d}
\end{eqnarray}
where $\beta=1/T$, $a=\beta\mu_{R}$. The two theta functions $\theta(k_{z})$
and $\theta(-k_{z})$ in $\ln\Xi$ is necessary, as discussed in \citep{Dong:2020zci}.
The vacuum terms in $\ln\Xi$ have been thrown away. The summations
for $k_{y}$ and $k_{z}$ in Eq. (\ref{eq:5.1d}) can be replaced
by the degeneracy factor $eBL^{2}/(2\pi)$ and the integral $(L/2\pi)\int dk_{z}$
respectively. Defining a dimensionless variable $b=2eB\beta^{2}$,
$\ln\Xi$ can be written as 
\begin{equation}
\ln\Xi=\frac{V}{\beta^{3}}g(a,b),\label{eq:5.1f}
\end{equation}
where $V=L^{3}$ and $g(a,b)$ is defined as
\begin{eqnarray}
g(a,b) & = & \frac{b}{8\pi^{2}}\int_{0}^{\infty}ds\bigg[\ln(1+e^{a-s})+\ln(1+e^{-a-s})\bigg]\nonumber \\
 &  & +\frac{b}{4\pi^{2}}\int_{0}^{\infty}ds\sum_{n=1}^{\infty}\bigg[\ln(1+e^{a-\sqrt{nb+s^{2}}})+\ln(1+e^{-a-\sqrt{nb+s^{2}}})\bigg].\label{eq:5.1g}
\end{eqnarray}

From the grand partition function $\ln\Xi$, the thermodynamic quantities
of the system, such as particle number $N=Vn$, energy $U=V\varepsilon$,
pressure $p$, entropy $S=Vs$ and magnetization intensity $M$, can
be expressed as
\begin{equation}
N=\frac{\partial}{\partial a}\ln\Xi,\label{eq:512e}
\end{equation}
\begin{equation}
U=-\frac{\partial}{\partial\beta}\ln\Xi,\label{eq:512f}
\end{equation}
\begin{equation}
p=\frac{1}{\beta}\frac{\partial}{\partial V}\ln\Xi,\label{eq:512g}
\end{equation}
\begin{equation}
S=\ln\Xi+\beta U-aN,\label{eq:512h}
\end{equation}
\begin{equation}
M=\frac{1}{\beta}\frac{\partial}{\partial B}\bigg(\frac{\ln\Xi}{V}\bigg).\label{eq:5.1h-1}
\end{equation}

Taking use of Eq. (\ref{eq:5.1f}), all intensive quantities, such
as particle number density $n$, energy density $\varepsilon$, pressure
$p$, entropy density $s$, magnetization intensity $M$, magnetic
susceptibility $\chi=\partial M/\partial B$ and heat capacity $c_{T}=\partial\varepsilon/\partial T$,
can be expressed by $g(a,b)$ as follows,
\begin{equation}
n=\frac{1}{\beta^{3}}\frac{\partial}{\partial a}g(a,b),\label{eq:512i}
\end{equation}
\begin{equation}
\varepsilon=\frac{1}{\beta^{4}}\bigg(3-2b\frac{\partial}{\partial b}\bigg)g(a,b),\label{eq:512j}
\end{equation}
\begin{equation}
p=\frac{1}{\beta^{4}}g(a,b),\label{eq:512k}
\end{equation}
\begin{equation}
s=\frac{1}{\beta^{3}}\bigg(4-a\frac{\partial}{\partial a}-2b\frac{\partial}{\partial b}\bigg)g(a,b),\label{eq:512l}
\end{equation}
\begin{equation}
M=\frac{2e}{\beta^{2}}\frac{\partial}{\partial b}g(a,b),\label{eq:512m}
\end{equation}
\begin{equation}
\chi=4e^{2}\frac{\partial^{2}}{\partial b^{2}}g(a,b),\label{eq:512n}
\end{equation}
\begin{equation}
c_{T}=\frac{1}{\beta^{3}}\bigg(12-3a\frac{\partial}{\partial a}-10b\frac{\partial}{\partial b}+2ab\frac{\partial^{2}}{\partial a\partial b}+4b^{2}\frac{\partial^{2}}{\partial b^{2}}\bigg)g(a,b).\label{eq:5.1i}
\end{equation}

\section{Expansions of intensive quantities with respect to $b$}

\label{sec:Expansion}To study the influence of the magnetic field
on the thermodynamics of the right-handed fermion system, in this
section we will expand all thermodynamic quantities as series with
respect to $b=2eB\beta^{2}$.

Defining an auxiliary function $f(a,x)$ as
\begin{equation}
f(a,x)=\ln(1+e^{a-x})+\ln(1+e^{-a-x}),\label{eq:a2}
\end{equation}
then $g(a,b)$ in Eq. (\ref{eq:5.1g}) becomes
\begin{equation}
g(a,b)=\frac{b}{4\pi^{2}}\int_{0}^{\infty}ds\left[\frac{1}{2}f(a,s)+\sum_{n=1}^{\infty}f(a,\sqrt{nb+s^{2}})\right].\label{eq:5.1j-1}
\end{equation}
In Appendix \ref{sec:Abel-Plana}, we have proven that, when $-\pi<a<\pi$,
the summation over Landau levels in the integrand in Eq. (\ref{eq:5.1j-1})
can be transformed into integrations by following Abel-Plana formula
\citep{Ni:2003,Butzer:2011},
\begin{equation}
\frac{1}{2}\mathcal{F}(0)+\sum_{n=1}^{\infty}\mathcal{F}(n)=\int_{0}^{\infty}dt\mathcal{F}(t)+i\int_{0}^{\infty}dt\frac{\mathcal{F}(it)-\mathcal{F}(-it)}{e^{2\pi t}-1}.\label{eq:5.1k-1}
\end{equation}
Then $g(a,b)$ can be expressed as
\begin{eqnarray}
g(a,b) & = & \bigg(\frac{7\pi^{2}}{360}+\frac{a^{2}}{12}+\frac{a^{4}}{24\pi^{2}}\bigg)\nonumber \\
 &  & +\frac{b}{4\pi^{2}}\times i\int_{0}^{\infty}ds\int_{0}^{\infty}dt\frac{f(a,\sqrt{itb+s^{2}})-f(a,\sqrt{-itb+s^{2}})}{e^{2\pi t}-1},\label{eq:5.1l-1}
\end{eqnarray}
where the first term comes from the first integration on the right-hand
side of Eq. (\ref{eq:5.1k-1}), 
\begin{equation}
\int_{0}^{\infty}ds\int_{0}^{\infty}dtf(a,\sqrt{t+s^{2}})=\frac{7\pi^{4}}{90}+\frac{\pi^{2}a^{2}}{3}+\frac{a^{4}}{6}.\label{eq:54o}
\end{equation}

In Appendix \ref{sec:Expansion-app}, we obtained the series expansion
of $g(a,b)$ at $b=0$ as follows,
\begin{eqnarray}
g(a,b) & = & \bigg(\frac{7\pi^{2}}{360}+\frac{a^{2}}{12}+\frac{a^{4}}{24\pi^{2}}\bigg)-\frac{b^{2}\ln b^{2}}{384\pi^{2}}\nonumber \\
 &  & -\frac{b^{2}}{96\pi^{2}}\ln\bigg(\frac{e^{1+C_{1}(a)}}{2G^{6}}\bigg)-\frac{1}{2\pi^{2}}\sum_{n=1}^{\infty}\frac{(4n+1)!!}{(4n+4)!!}\mathscr{B}_{2n+2}C_{2n+1}(a)b^{2n+2},\label{eq:5.1m}
\end{eqnarray}
where $G=1.282427$ is the Glaisher number, $\mathscr{B}_{n}$ are
Bernoulli numbers, and $C_{2n+1}(a)\ (n\geqslant0)$ is 
\begin{equation}
C_{2n+1}(a)=-\delta_{n,0}+\frac{1}{(4n+1)!}\int_{0}^{\infty}dy\ln y\frac{d^{4n+1}}{dy^{4n+1}}\left(\frac{1}{e^{y+a}+1}+\frac{1}{e^{y-a}+1}\right).\label{eq:5.1n}
\end{equation}

In the series of $g(a,b)$, besides the terms with $b^{2n}$, there
is also a single singular term $b^{2}\ln b^{2}$, which indicates
that $g(a,b)$ is not analytic at $b=0$. When $a=0$, as calculated
in Appendix \ref{sec:C(2n+1)at-0}, the integration in Eq. (\ref{eq:5.1n})
can be analytically integrated out, 
\begin{equation}
C_{2n+1}(0)=(\ln4+\gamma-1)\delta_{n,0}+\frac{2\zeta^{\prime}(-4n)}{(4n+1)!}\left(2^{4n+1}-1\right).\label{eq:518a}
\end{equation}

Taking use of Eqs. (\ref{eq:512i})-(\ref{eq:5.1i}), all intensive
quantities, such as particle number density $n$, energy density $\varepsilon$,
pressure $p$, entropy density $s$, magnetization intensity $M$,
magnetic susceptibility $\chi$ and heat capacity $c_{T}$, can be
expressed as series of $b$ at $b=0$ in the following,
\begin{equation}
n\beta^{3}=\left(\frac{a}{6}+\frac{a^{3}}{6\pi^{2}}\right)-\frac{1}{2\pi^{2}}\sum_{n=0}^{\infty}\frac{(4n+1)!!}{(4n+4)!!}\mathscr{B}_{2n+2}C_{2n+1}^{\prime}(a)b^{2n+2},\label{eq:54p}
\end{equation}

\begin{eqnarray}
\varepsilon\beta^{4} & = & \left(\frac{7\pi^{2}}{120}+\frac{a^{2}}{4}+\frac{a^{4}}{8\pi^{2}}\right)+\frac{b^{2}\ln b^{2}}{384\pi^{2}}+\frac{b^{2}}{96\pi^{2}}\ln\left(\frac{e^{2+C_{1}(a)}}{2G^{6}}\right)\nonumber \\
 &  & +\frac{1}{2\pi^{2}}\sum_{n=1}^{\infty}\frac{(4n+1)!!}{(4n+4)!!}(4n+1)\mathscr{B}_{2n+2}C_{2n+1}(a)b^{2n+2},\label{eq:a11}
\end{eqnarray}
\begin{eqnarray}
p\beta^{4} & = & \left(\frac{7\pi^{2}}{360}+\frac{a^{2}}{12}+\frac{a^{4}}{24\pi^{2}}\right)-\frac{b^{2}\ln b^{2}}{384\pi^{2}}-\frac{b^{2}}{96\pi^{2}}\ln\left(\frac{e^{1+C_{1}(a)}}{2G^{6}}\right)\nonumber \\
 &  & -\frac{1}{2\pi^{2}}\sum_{n=1}^{\infty}\frac{(4n+1)!!}{(4n+4)!!}\mathscr{B}_{2n+2}C_{2n+1}(a)b^{2n+2},\label{eq:54q}
\end{eqnarray}

\begin{eqnarray}
s\beta^{3} & = & \left(\frac{7\pi^{2}}{90}+\frac{a^{2}}{6}\right)+\frac{1+aC_{1}^{\prime}(a)}{96\pi^{2}}b^{2}\nonumber \\
 &  & +\frac{1}{2\pi^{2}}\sum_{n=1}^{\infty}\frac{(4n+1)!!}{(4n+4)!!}\mathscr{B}_{2n+2}\left(4n+a\frac{d}{da}\right)C_{2n+1}(a)b^{2n+2},\label{eq:54r}
\end{eqnarray}
\begin{eqnarray}
M\beta^{2}/e & = & -\frac{b\ln b^{2}}{96\pi^{2}}-\frac{b}{24\pi^{2}}\ln\bigg(\frac{e^{5/4+C_{1}(a)}}{2G^{6}}\bigg)\nonumber \\
 &  & -\frac{1}{2\pi^{2}}\sum_{n=1}^{\infty}\frac{(4n+1)!!}{(4n+2)!!}\mathscr{B}_{2n+2}C_{2n+1}(a)b^{2n+1},\label{eq:54s}
\end{eqnarray}
\begin{eqnarray}
\chi/e^{2} & = & -\frac{\ln b^{2}}{48\pi^{2}}-\frac{1}{12\pi^{2}}\ln\bigg(\frac{e^{7/4+C_{1}(a)}}{2G^{6}}\bigg)\nonumber \\
 &  & -\frac{1}{2\pi^{2}}\sum_{n=1}^{\infty}\frac{(4n+1)!!}{(4n)!!}\mathscr{B}_{2n+2}C_{2n+1}(a)b^{2n},\label{eq:54t}
\end{eqnarray}
\begin{eqnarray}
c_{T}\beta^{3} & = & \left(\frac{7\pi^{2}}{30}+\frac{a^{2}}{2}\right)-\frac{1+aC_{1}^{\prime}(a)}{96\pi^{2}}b^{2}\nonumber \\
 &  & -\frac{1}{2\pi^{2}}\sum_{n=1}^{\infty}\frac{(4n+1)!!}{(4n+4)!!}(4n+1)\mathscr{B}_{2n+2}\left(4n+a\frac{d}{da}\right)C_{2n+1}(a)b^{2n+2},\label{eq:54u}
\end{eqnarray}
where we have used the temperature factor $\beta$ and the electric
charge $e$ to cancle the dimensions of all intensive quantities.
We can see that in the expansions of energy density $\varepsilon$,
pressure $p$, magnetization intensity $M$ and magnetic susceptibility
$\chi$, besides the terms with $b^{n}$, there is also a term involving
$\ln b^{2}$, which is singular at $b=0$. Nevertheless, the expansions
of particle number density $n$, entropy density $s$ and heat capacity
$c_{T}$ are just power series of $b^{2}$, i.e. they are analytic
at $b=0$. 

For left-handed fermion system, we can simply replace $a=\beta\mu_{R}$
with $\beta\mu_{L}$ in the expressions obtained above for right-handed
fermion system, and all intensive quantities of chiral fermion system
can be obtained as a summation of the results of left-handed and right-handed
fermion systems.

\section{Effect of magnetic field on thermodynamic quantities}

\label{sec:-extreme}In this section, we will investigate the effect
of magnetic field on the thermodynamics of the chiral fermion system.
Since the dimensionless variables $a=\mu_{R}/T$ and $b=2eB/T^{2}$
are proportional to the chemical potential $\mu_{R}$ and the magnetic
field $B$ respectively, we will often use $a$ and $b$ instead of
$\mu_{R}$ and $B$ to represent the chemical potential and the magnetic
field in following discussion. For a fixed chemical potential $a$,
we take $Q(a,b)$ as the thermodynamic quantity with nonzero $b$
and $Q(a,0)$ as the one with $b=0$.

According to Eqs. (\ref{eq:512i})-(\ref{eq:5.1i}) and the integration
expression of $g(a,b)$ in Eq. (\ref{eq:5.1l-1}), we can plot the
curves of all thermodynamic quantities as functions of $b$ and $a$.
The asymptotic behaviors of all thermodynamic quantities as $b\rightarrow0$
can be obtained directly by Eqs. (\ref{eq:54p})-(\ref{eq:54u}).
For $b\rightarrow\infty$ limit, we must take use of following asymptotic
formula of $g(a,b)$, as calculated in Appendix \ref{sec:asymptotic},

\begin{equation}
\lim_{b\rightarrow\infty}g(a,b)=\frac{b}{48\pi^{2}}(\pi^{2}+3a^{2}).\label{eq:54c}
\end{equation}
Then Eqs. (\ref{eq:512i})-(\ref{eq:5.1i}) can give the leading order
term for all thermodynamic quantities as $b\rightarrow\infty$.

\subsection{Particle number density}

The asymptotic behavior of particle number density ratio $n(a,b)/n(a,0)$
as $b\rightarrow0$ is
\begin{equation}
\frac{n(a,b)}{n(a,0)}\rightarrow1-\frac{C_{1}^{\prime}(a)b^{2}}{16(\pi^{2}a+a^{3})}.\label{eq:54g}
\end{equation}
Taking use of Eqs. (\ref{eq:512o}) and (\ref{eq:510b5}) in the appendix,
the coefficient of $b^{2}$ in Eq. (\ref{eq:54g}) tends to be $-7\zeta^{\prime}(-2)/(8\pi^{2})=0.0027$
as $a\rightarrow0$ and tends to be zero as $a\rightarrow\infty$,
which indicates that in weak magnetic field limit the enhancement
of particle number density is smaller for larger chemical potential.
It is worth pointing out that Eq. (\ref{eq:54g}) is consistent with
the second order result of particle number density in weak electromagnetic
field expansion by some of us in \citep{Yang:2020mtz}, where the
Wigner function approach is used. 

The asymptotic behavior of $n(a,b)/n(a,0)$ as $b\rightarrow\infty$
is

\begin{equation}
\frac{n(a,b)}{n(a,0)}\rightarrow\frac{3b}{4(\pi^{2}+a^{2})}\label{eq:55a}
\end{equation}
which increases linearly as $b\rightarrow\infty$ and increases more
slowly for larger $a$. 

The asymptotic behavior of $n(a,b)/n(a,0)$ as $a\rightarrow0$ is
listed as follows,
\begin{equation}
\lim_{a\rightarrow0}\frac{n(a,b)}{n(a,0)}=1+\frac{3b}{\pi^{2}}\times i\int_{0}^{\infty}ds\int_{0}^{\infty}dt\frac{1}{e^{2\pi t}-1}\left[\frac{e^{\sqrt{itb+s^{2}}}}{(1+e^{\sqrt{itb+s^{2}}})^{2}}-\frac{e^{\sqrt{-itb+s^{2}}}}{(1+e^{\sqrt{-itb+s^{2}}})^{2}}\right].\label{eq:510g1}
\end{equation}

\begin{figure}[H]
\begin{centering}
\includegraphics[scale=0.55]{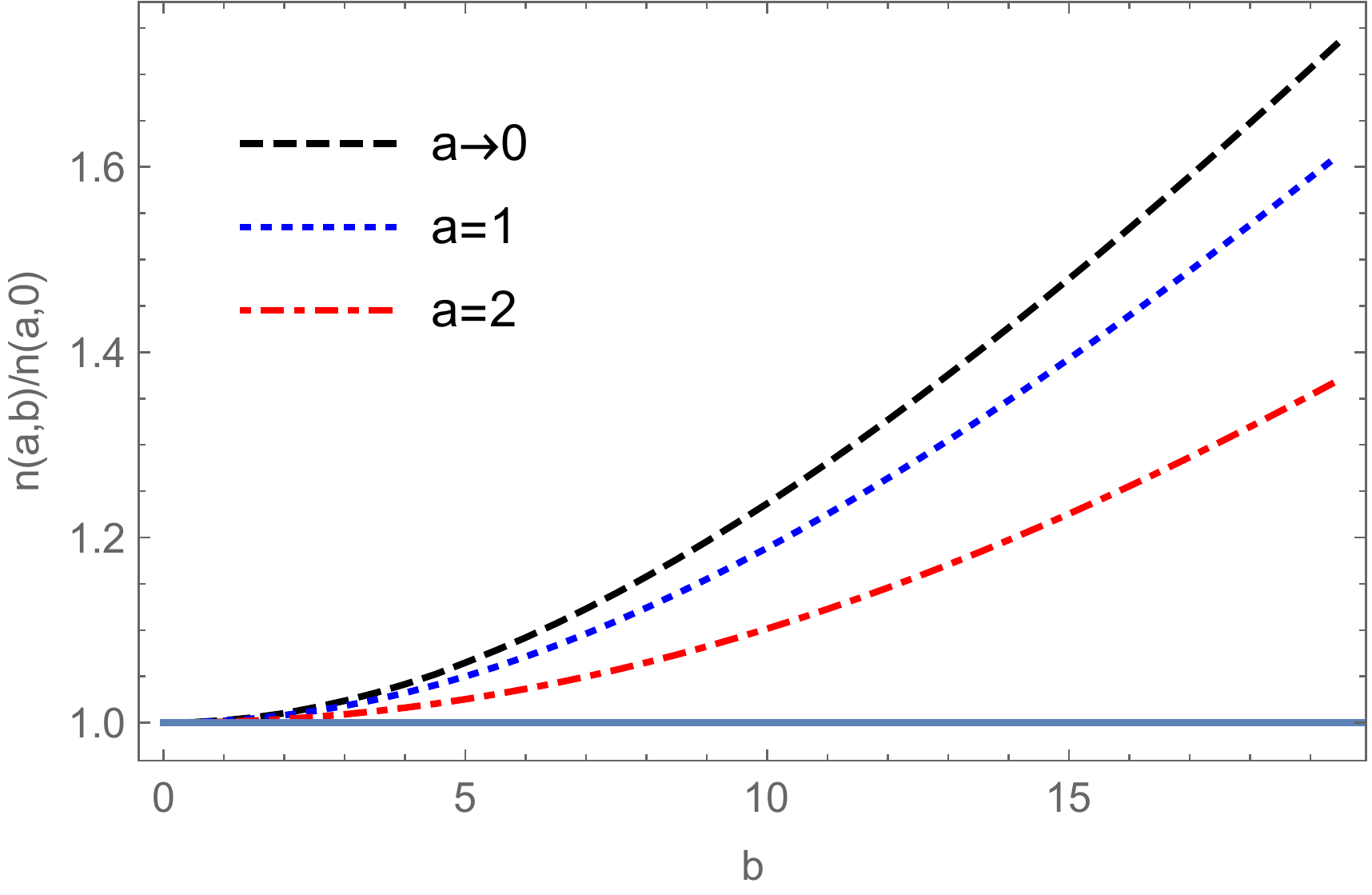}
\par\end{centering}
\caption{\label{fig:number density}Curves of particle number density ratio
$n(a,b)/n(a,0)$ with respect to $b$ for $a=1,2$ and $a\rightarrow0$}

\end{figure}

In Figure \ref{fig:number density}, we plot the curves of $n(a,b)/n(a,0)$
with respect to $b$ for $a=1,2$ and $a\rightarrow0$. We can see
that the existence of magnetic field can considerably enhance the
particle number density, and the enhancement is larger when $b$ is
stronger and $a$ is smaller. The trends of the curves are consistent
with our asymptotic analysis.

\subsection{Energy density and heat capacity}

From the expressions of the grand partition function $\ln\Xi$ and
energy $U$ in Eqs. (\ref{eq:5.1d}) and (\ref{eq:512f}), we can
express the energy density $\varepsilon$ by following form
\begin{eqnarray}
\varepsilon & = & \frac{1}{V}\sum_{k_{y},k_{z}}\bigg(\frac{k_{z}\theta(k_{z})}{e^{\beta(k_{z}-\mu_{R})}+1}+\frac{(-k_{z})\theta(-k_{z})}{e^{\beta(-k_{z}+\mu_{R})}+1}\bigg)\nonumber \\
 &  & +\frac{1}{V}\sum_{n=1}^{\infty}\sum_{k_{y},k_{z}}E_{n}(k_{z})\bigg[\frac{1}{e^{\beta[E(n,k_{z})-\mu_{R}]}+1}+\frac{1}{e^{\beta[E(n,k_{z})+\mu_{R}]}+1}\bigg],\label{eq:54a}
\end{eqnarray}
The physical meaning of Eq. (\ref{eq:54a}) is very clear: the whole
energy of the system is jointly determined by the energy and the particle
number of every quantum state and the degeneracy factor $eBL^{2}/(2\pi)$
of every Landau level. As the magnetic field increases, the particle
number of every state will decrease rapidly as $e^{-\sqrt{b}}$ due
to Fermi-Dirac distribution, meanwhile the product of the energy and
the degeneracy of every Landau level increases as $b^{3/2}$, which
indicates the existence of an extremum of energy density as $b$ increases.

\begin{figure}[H]
\begin{centering}
\includegraphics[scale=0.55]{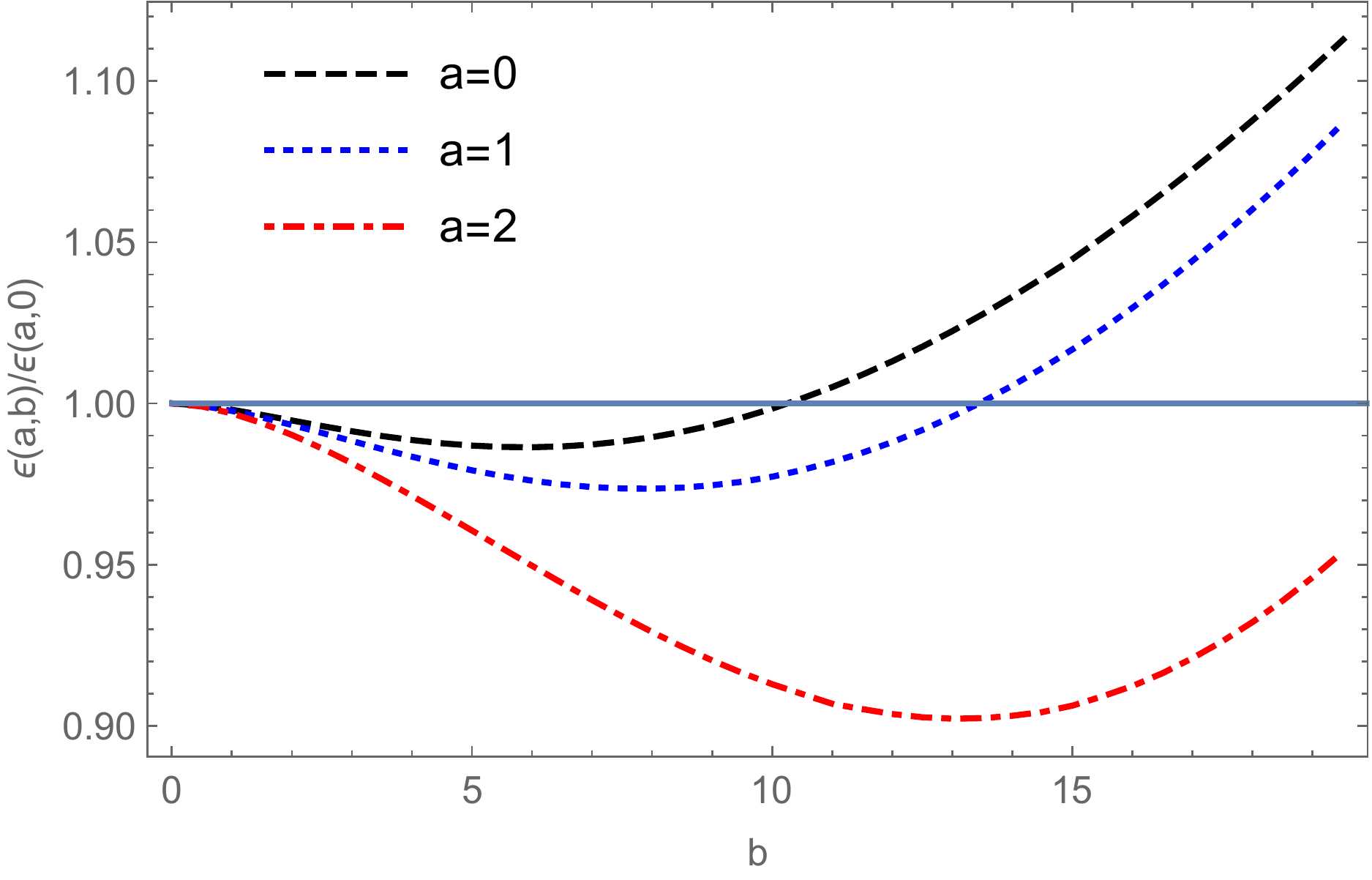}
\par\end{centering}
\caption{\label{fig:energy}Curves of energy density ratio $\varepsilon(a,b)/\varepsilon(a,0)$
with respect to $b$ for $a=0,1,2$}

\end{figure}

In Figure \ref{fig:energy}, we plot the curves of energy density
ratio $\varepsilon(a,b)/\varepsilon(a,0)$ with respect to $b$ for
$a=0,1,2$. We can see that the curves decline first and then rise,
leading to the appearance of extremum as expected. When $a=0$, the
extremum locates at about $(5.89,0.9864)$, and $\varepsilon(a,b)/\varepsilon(a,0)=1$
at $b\approx10.25$. When $a$ increases, the extremum moves to the
lower right corner. 

The asymptotic behavior of $\varepsilon(a,b)/\varepsilon(a,0)$ as
$b\rightarrow0$ is
\begin{equation}
\frac{\varepsilon(a,b)}{\varepsilon(a,0)}\rightarrow1+\frac{5b^{2}\ln b^{2}}{16(7\pi^{4}+30\pi^{2}a^{2}+15a^{4})}.\label{eq:513i}
\end{equation}
Since $b^{2}\ln b^{2}<0$ as $b<1$, the energy density in weak magnetic
field limit is smaller than the one without magnetic field. 

The asymptotic behavior of $\varepsilon(a,b)$ as $b\rightarrow\infty$
is
\begin{equation}
\frac{\varepsilon(a,b)}{\varepsilon(a,0)}\rightarrow\frac{5\pi^{2}+15a^{2}}{14\pi^{4}+60\pi^{2}a^{2}+30a^{4}}\times b,\label{eq:513o}
\end{equation}
which increases linearly as $b\rightarrow\infty$ and increases more
slowly for larger $a$. The trend of the curves in Figure \ref{fig:energy}
are consistent with our asymptotic analysis.

If we take high temperature limit, i.e. $b\ll1$ and $a\ll1$, then
the energy density becomes
\begin{equation}
\varepsilon=\frac{7\pi^{2}}{120}T^{4}+\frac{e^{2}B^{2}}{96\pi^{2}}\ln\left(\frac{e^{2}B^{2}}{T^{4}}\right).\label{eq:513j}
\end{equation}
We emphasize that the $\ln T$ term for high temperature limit in
Eq. (\ref{eq:513j}) is also consistent with the second order result
of energy density in \citep{Yang:2020mtz}. 

If we take low temperature limit and $a=0$, then the energy density
becomes
\begin{equation}
\varepsilon=\left\{ \begin{array}{cc}
\frac{1}{24}eBT^{2}, & B\neq0\\
\frac{7\pi^{2}}{120}T^{4}, & B=0
\end{array}\right.,\label{eq:513j-1}
\end{equation}
which decreases to be zero as $T\rightarrow0$. We can see that the
existence of magnetic field makes the asymptotic behavior of the energy
density become $T^{2}$ instead of $T^{4}$ as $T\rightarrow0$.

\begin{figure}[H]
\begin{centering}
\includegraphics[scale=0.6]{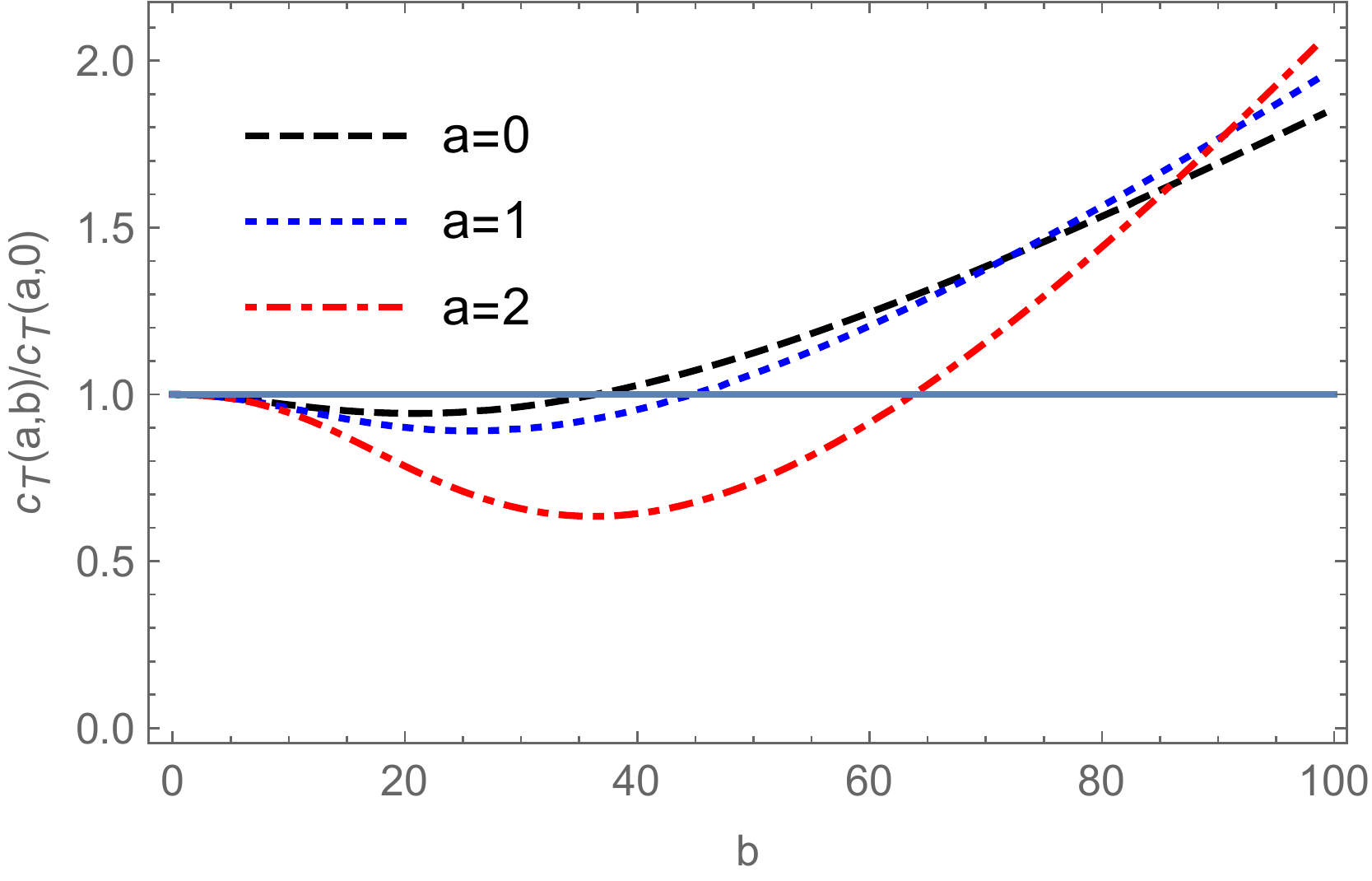}
\par\end{centering}
\caption{\label{fig:heat}Curves of heat capacity ratio $c_{T}(a,b)/c_{T}(a,0)$
with respect to $b$ for $a=0,1,2$}

\end{figure}

In Figure \ref{fig:heat}, we plot the curves of heat capacity ratio
$c_{T}(a,b)/c_{T}(a,0)$ with respect to $b$ for $a=0,1,2$, where
there are also extremum. For high temperature limit, the heat capacity
becomes 
\begin{equation}
c_{T}=\frac{7\pi^{2}}{30}T^{3}-\frac{e^{2}B^{2}}{24\pi^{2}T}.\label{eq:510e}
\end{equation}
 For low temperature limit with $a=0$, the heat capacity becomes
\begin{equation}
c_{T}=\left\{ \begin{array}{cc}
\frac{1}{12}eBT, & B\neq0\\
\frac{7\pi^{2}}{30}T^{3}, & B=0
\end{array}\right.,\label{eq:513p}
\end{equation}
which tends to be zero as $T\rightarrow0$. Similar to the case of
energy density, the existence of magnetic field makes the asymptotic
behavior of the heat capacity become $T$ instead of $T^{3}$ as $T\rightarrow0$. 

\subsection{Pressure and entropy density}

\begin{figure}[H]
\begin{centering}
\includegraphics[scale=0.55]{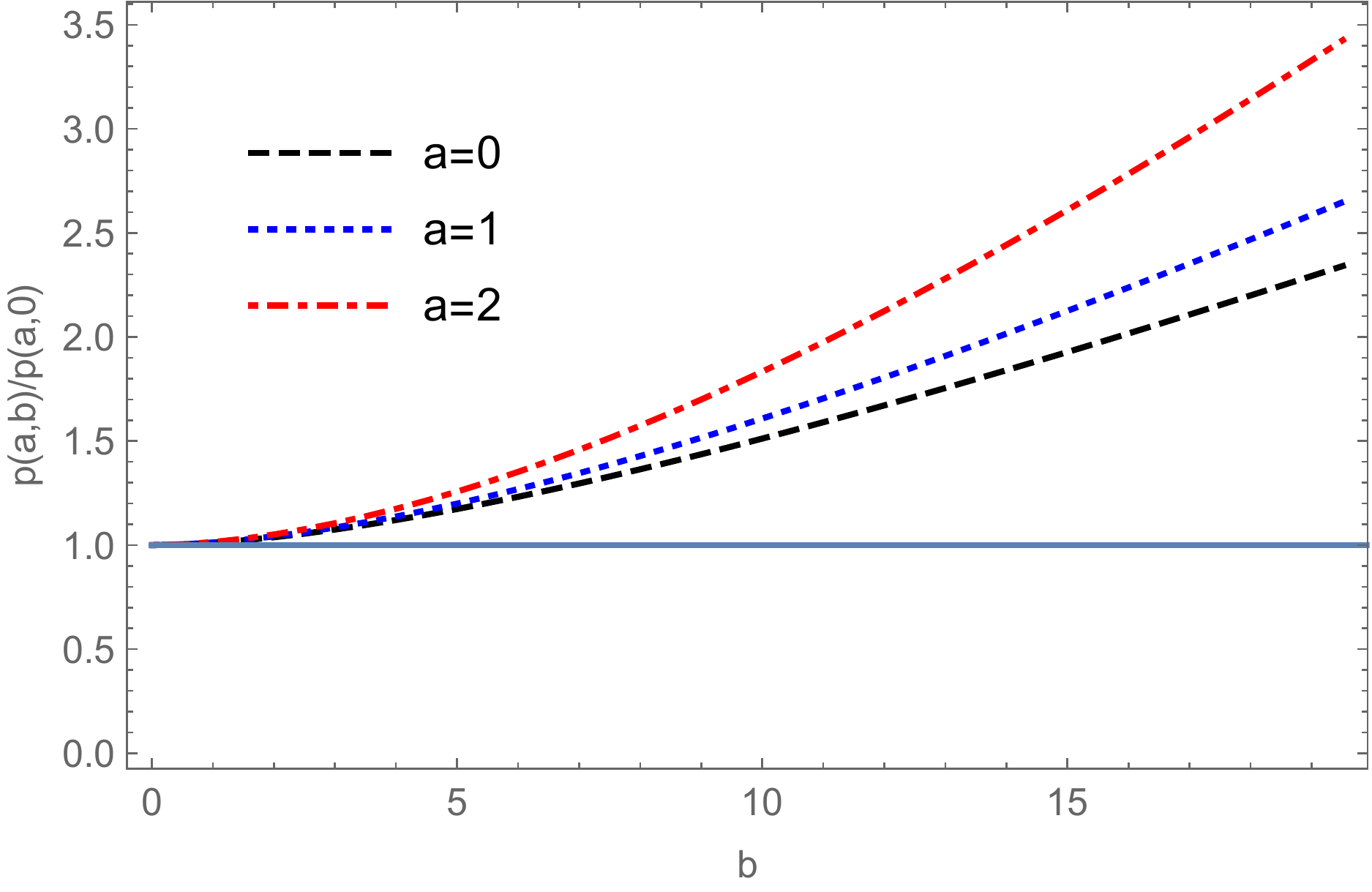}
\par\end{centering}
\caption{\label{fig:pressure}Curves of pressure ratio $p(a,b)/p(a,0)$ with
respect to $b$ for $a=0,1,2$}

\end{figure}

\begin{figure}[H]
\begin{centering}
\includegraphics[scale=0.55]{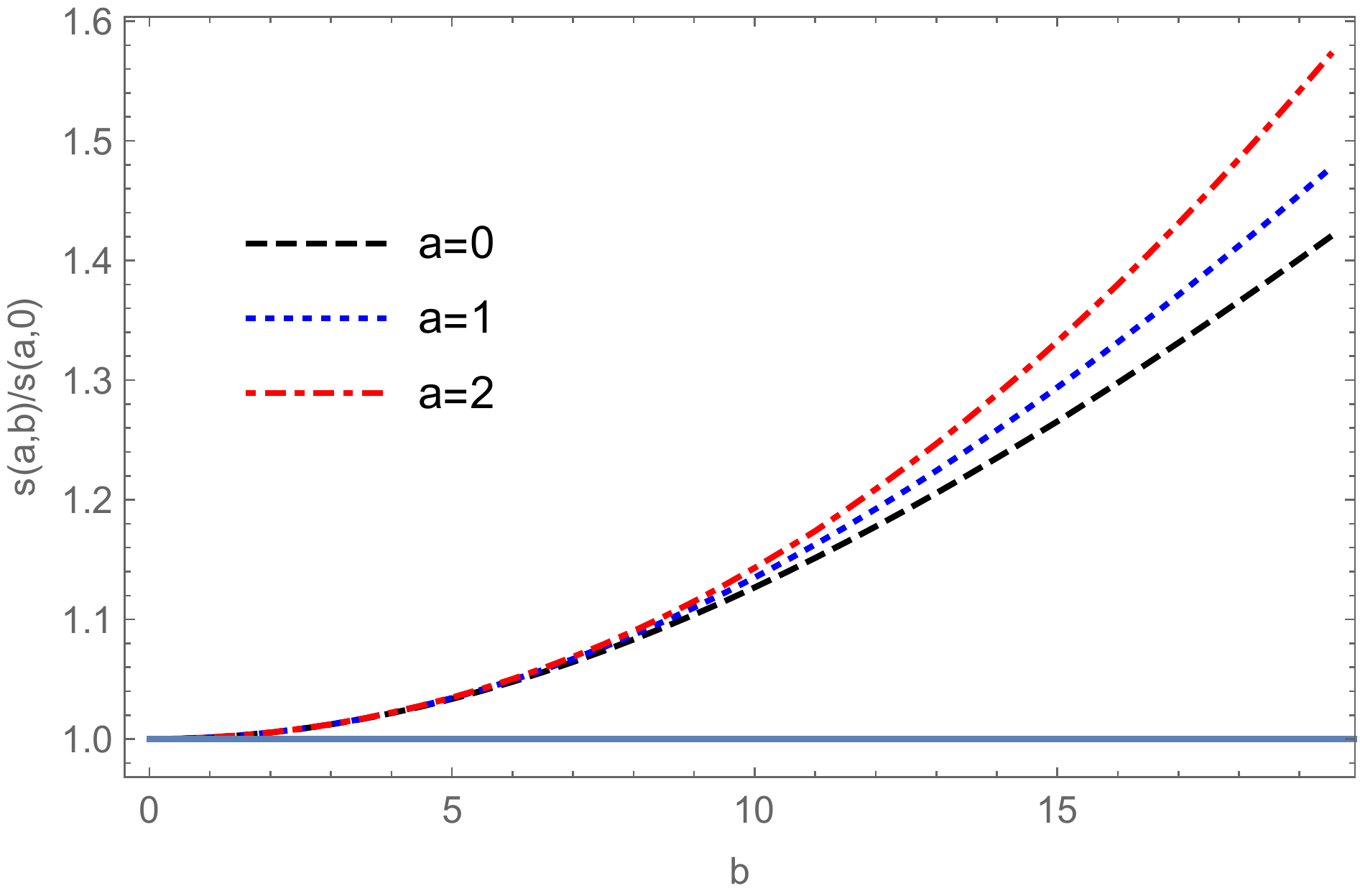}
\par\end{centering}
\caption{\label{fig:entropy}Curves of entropy density ratio $s(a,b)/s(a,0)$
with respect to $b$ for $a=0,1,2$}

\end{figure}

In Figures \ref{fig:pressure} and \ref{fig:entropy}, we plot the
curves of pressure ratio $p(a,b)/p(a,0)$ and entropy density ratio
$s(a,b)/s(a,0)$ with respect to $b$ for $a=0,1,2$, which are both
increasing functions of $b$, i.e. the existence of the magnetic field
can enhance the pressure and entropy density of the system. 

For high temperature limit, the pressure and entropy density become
\begin{equation}
p=\frac{7\pi^{2}}{360}T^{4}-\frac{e^{2}B^{2}}{96\pi^{2}}\ln\left(\frac{e^{2}B^{2}}{T^{4}}\right),\label{eq:510g}
\end{equation}
\begin{equation}
s=\frac{7\pi^{2}}{90}T^{3}+\frac{e^{2}B^{2}}{24\pi^{2}T}.\label{eq:510h}
\end{equation}
For low temperature limit with $a=0$, the pressure and entropy density
become
\begin{equation}
p=\left\{ \begin{array}{cc}
\frac{1}{24}eBT^{2}, & B\neq0\\
\frac{7\pi^{2}}{360}T^{4}, & B=0
\end{array}\right.,\label{eq:513k}
\end{equation}

\begin{equation}
s=\left\{ \begin{array}{cc}
\frac{1}{12}eBT, & B\neq0\\
\frac{7\pi^{2}}{90}T^{3}, & B=0
\end{array}\right.,\label{eq:513l}
\end{equation}
which both tend to be zero as $T\rightarrow0$. Similar to the cases
of energy density and heat capacity, the asymptotic behaviors of the
pressure and the entropy density as $T\rightarrow0$ are also changed
due to the existence of magnetic field.

\subsection{Magnetization intensity and magnetic susceptibility }

\begin{figure}[H]
\begin{centering}
\includegraphics[scale=0.55]{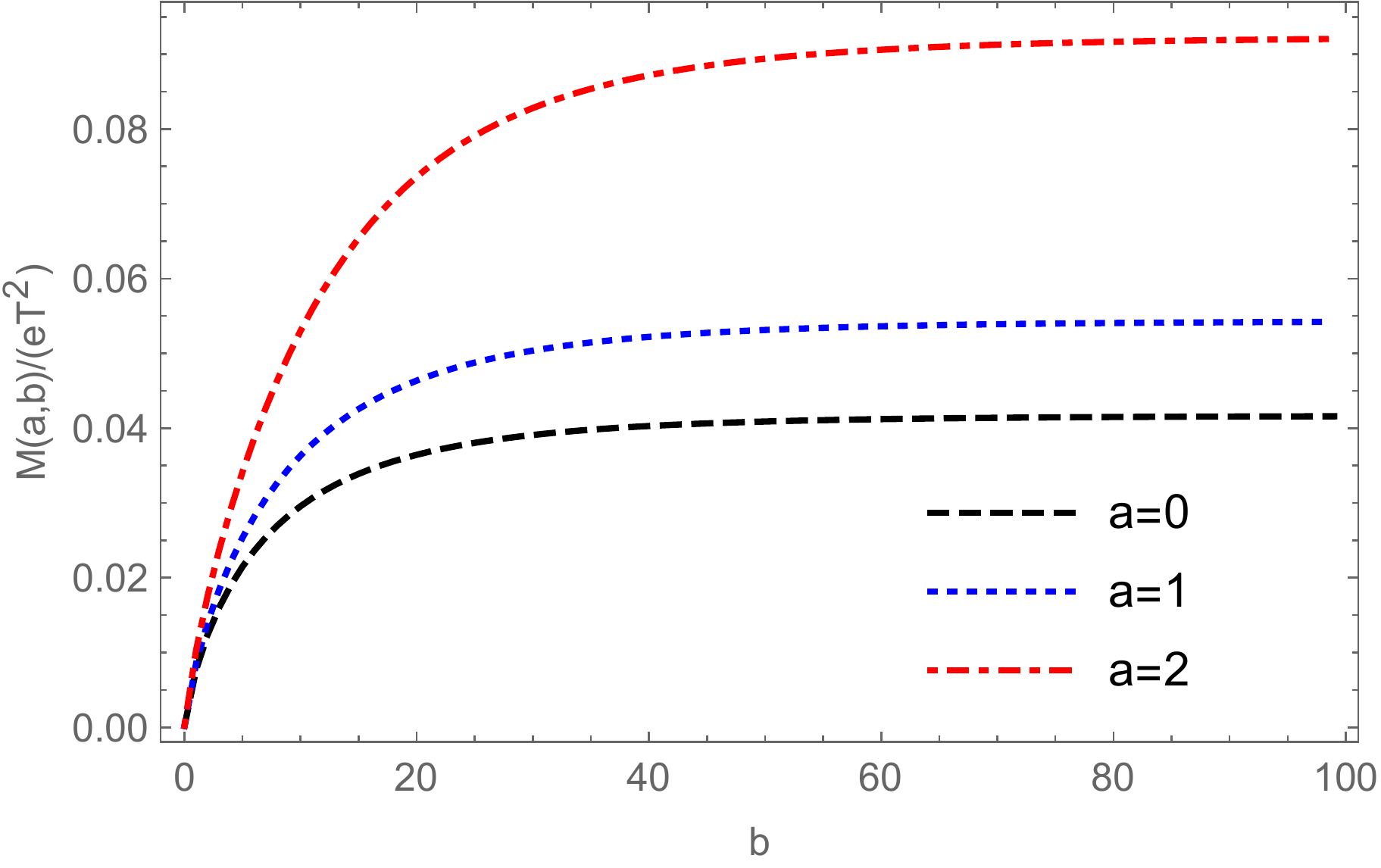}
\par\end{centering}
\caption{\label{fig:magnetization}Curves of magnetization intensity $M(a,b)/(eT^{2})$
with respect to $b$ for $a=0,1,2$}
\end{figure}

In Figure \ref{fig:magnetization}, we plot the curves of magnetization
intensity $M(a,b)$ with respect to $b$ for $a=0,1,2$. As $b$ increases,
the magnetization intensity increases rapidly from zero to saturation.
The saturation value $M_{0}$ depends on the chemical potential as
follows,
\begin{equation}
M_{0}(a)=\frac{eT^{2}}{24}\left(1+\frac{3a^{2}}{\pi^{2}}\right).\label{eq:510k}
\end{equation}
Then we have $M_{0}(0)=0.042\,eT^{2}$, $M_{0}(1)=0.054\,eT^{2}$,
$M_{0}(2)=0.092\,eT^{2}$, which are consistent with the saturation
values in Figure \ref{fig:magnetization}.

\begin{figure}[H]
\begin{centering}
\includegraphics[scale=0.55]{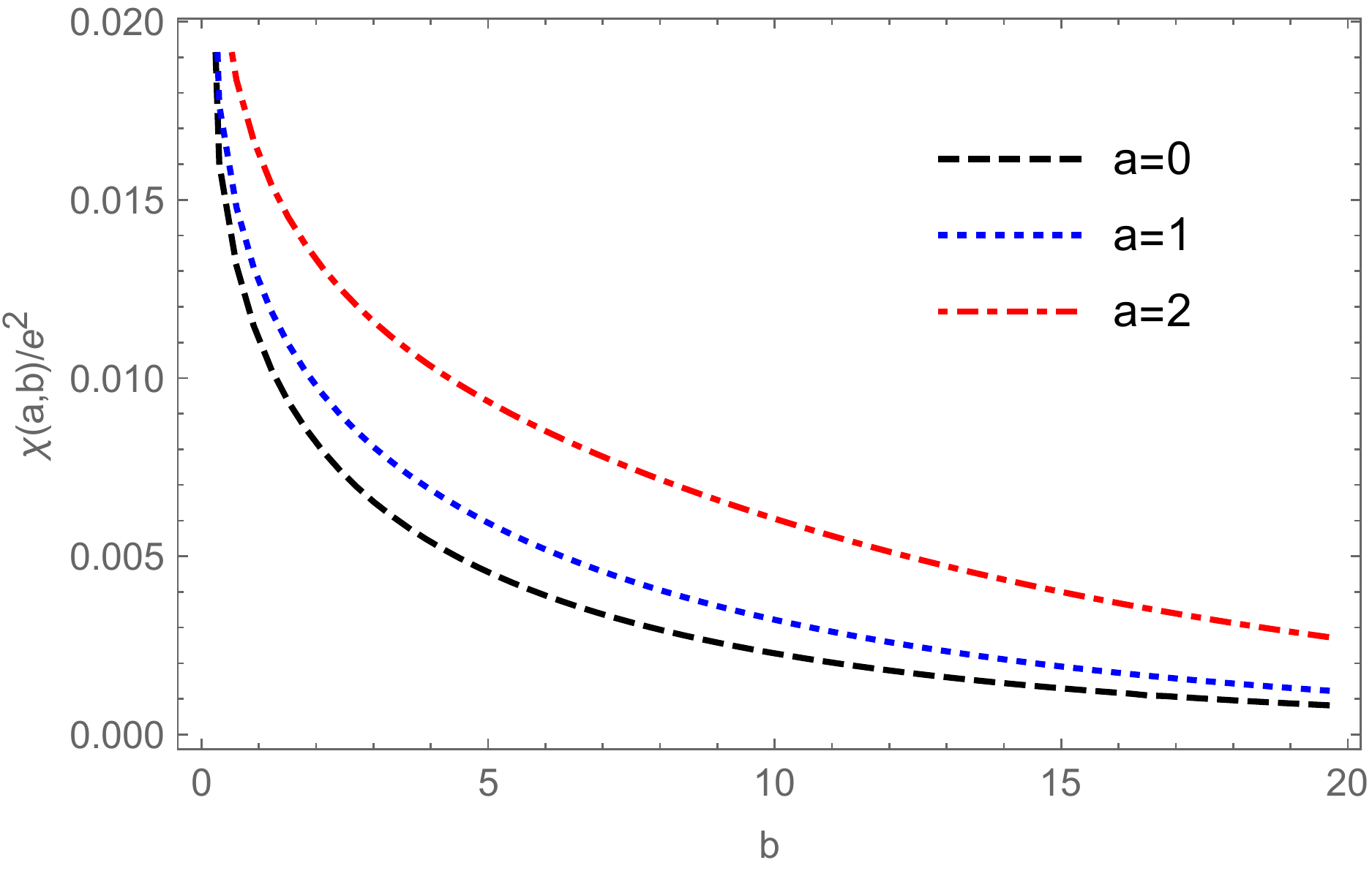}
\par\end{centering}
\caption{\label{fig:sus}Curves of magnetic susceptibility $\chi(a,b)/e^{2}$
with respect to $b$ for $a=0,1,2$}

\end{figure}

Since the magnetization intensity $M(a,b)$ becomes saturation as
$b\rightarrow\infty$ , the magnetic susceptibility $\chi(a,b)$ tends
to be zero as $b\rightarrow\infty$. In Figure \ref{fig:sus}, we
plot the curves of magnetic susceptibility $\chi(a,b)$ with respect
to $b$ for $a=0,1,2$, where the curves tend to be zero as $b\rightarrow\infty$
and tend to be divergent as $b\rightarrow0$. In fact, the magnetic
susceptibility in weak magnetic field limit or high temperature limit
becomes 
\begin{equation}
\chi=-\frac{e^{2}}{48\pi^{2}}\ln\left(\frac{e^{2}B^{2}}{T^{4}}\right),\label{eq:510o}
\end{equation}
which is logarithmically divergent as $B\rightarrow0$ or $T\rightarrow\infty$.
The $\ln T$ term in Eq. (\ref{eq:510o}) is also derived in a recent
article \citep{Bali:2020bcn}, where the authors have calculated the
high temperature expansion of the magnetic susceptibility of QCD matter
with physical quark mass and the leading order term is just $\ln T$
with the same coefficient as ours. 

In low temperature limit, i.e. $b\rightarrow\infty$, the magnetic
susceptibility tends to be zero, indicating that the system reaches
the magnetic saturation state. This is consistent with the numerical
result of lattice method for QCD matter system \citep{Bonati:2013qra,Bali:2020bcn}.

In Figure \ref{fig:sus}, we can see that $\chi$ is always positive,
so the chiral fermion system is a paramagnetic system.

\subsection{A brief summary }

We have studied the effect of magnetic field on the thermodynamics
of the right-handed fermion system. All results calculated above can
be directly generalised to the left-handed case. In Table \ref{tab:Leading-order-1},
we list the leading order terms of all intensive quantities of chiral
fermion system as $b\rightarrow0$ and $b\rightarrow\infty$, with
$a=0$.

\begin{table}[H]
\begin{centering}
\begin{tabular}{|c|c|c|c|c|c|c|c|c|}
\hline 
Quantities & $n$ & $\varepsilon\times\beta^{4}$ & $p\times\beta^{4}$ & $s\times\beta^{3}$ & $M\times\beta^{2}$ & $\chi$ & $c_{T}\times\beta^{3}$ & $T^{11}\times\beta^{4}$\tabularnewline
\hline 
\hline 
$b\rightarrow0$ & $0$ & $\frac{7\pi^{2}}{60}+\frac{b^{2}\ln b^{2}}{192\pi^{2}}$ & $\frac{7\pi^{2}}{180}-\frac{b^{2}\ln b^{2}}{192\pi^{2}}$ & $\frac{7\pi^{2}}{45}+\frac{b^{2}}{48\pi^{2}}$ & $-\frac{eb\ln b^{2}}{48\pi^{2}}$ & $-\frac{e^{2}\ln b^{2}}{24\pi^{2}}$ & $\frac{7\pi^{2}}{15}-\frac{b^{2}}{48\pi^{2}}$ & $\frac{7\pi^{2}}{180}+\frac{b^{2}\ln b^{2}}{192\pi^{2}}$\tabularnewline
\hline 
$b\rightarrow\infty$ & $0$ & $\frac{b}{24}$ & $\frac{b}{24}$ & $\frac{b}{12}$ & $\frac{e}{12}$ & $0$ & $\frac{b}{12}$ & $0$\tabularnewline
\hline 
\end{tabular}
\par\end{centering}
\caption{\label{tab:Leading-order-1}Leading order terms of all intensive quantities
of chiral fermion system as $b\rightarrow0$ and $b\rightarrow\infty$
(with $a=0$).}
\end{table}

In summary, we conclude that the magnetic field has great influence
on the thermodynamics of chiral fermion system. In high energy heavy
ion collisions, a new matter state ``Quark Gluon Plasma'' can be
created \citep{Arsene:2004fa,Adams:2005dq,Adcox:2004mh}, where the
quarks and gluons are deconfined from the interior of the hadrons
due to the high temperature and the chiral symmetry may be recovered
\citep{Ishihara:1999jn,Ishihara:1999ts}. For peripheral collisions,
there will produce huge magnetic field in the collision region \citep{Skokov:2009qp,Bzdak:2011yy,Voronyuk:2011jd,Deng:2012pc},
which may considerably change the properties of the deconfined matter
state and hence influence the distributions in phase space of the
final particles from the collision.

\section{Energy-momentum tensor}

\label{sec:tensor}In this section, we will show that how the nonzero
components of $T^{\mu\nu}$ of the right-handed fermion system are
related to $g(a,b)$. The $00$-component of $T^{\mu\nu}$ is just
the energy density $\varepsilon$, which has been calculated through
grand partition function in Secs. (\ref{sec:partition function})
and (\ref{sec:Expansion}). Now we want to calculate other components
of $T^{\mu\nu}$. 

Since the thermodynamic system is located in a uniform magnetic field
$\boldsymbol{B}=B\boldsymbol{e}_{z}$, all thermodynamic quantities
will be unchanged under the rotation along $z$-axis. The rotation
by an angle $\phi$ along $z$-axis can be represented by following
matrix,
\begin{equation}
\Lambda_{\ \nu}^{\mu}(\phi)=\left(\begin{array}{cccc}
1 & 0 & 0 & 0\\
0 & \cos\phi & -\sin\phi & 0\\
0 & \sin\phi & \cos\phi & 0\\
0 & 0 & 0 & 1
\end{array}\right).\label{eq:510p}
\end{equation}
In order to keep $T^{\mu\nu}$ unchanged under the rotation matrix
$\Lambda_{\ \nu}^{\mu}(\phi)$ for any angle $\phi$, i.e. $T^{\mu\nu}=\Lambda_{\ \alpha}^{\mu}\Lambda_{\ \beta}^{\nu}T^{\alpha\beta}$,
the components of $T^{\mu\nu}$ must satisfy these conditions,
\begin{equation}
T^{01}=T^{02}=T^{12}=T^{13}=T^{23}=0,\ T^{11}=T^{22},\label{eq:510q}
\end{equation}
i.e. the nonzero components are $T^{00}$, $T^{33}$, $T^{11}=T^{22}$,
$T^{03}$. According to Eq. (\ref{eq:5.1c}), the trace of $T^{\mu\nu}$
is zero, so the independent components are $T^{00},T^{11},T^{03}$. 

We can not directly calculate $T^{11}$ and $T^{03}$ by the grand
partition function. Alternatively, we appeal to ensemble average approach,
in which all macroscopic quantities are the ensemble average of normal
ordering of the corresponding operators. The symmetric and gauge invariant
energy-momentum tensor is 
\begin{equation}
T^{\mu\nu}=\frac{1}{2}\left\langle :\psi_{R}^{\dagger}i\sigma^{\mu}D^{\nu}\psi_{R}+\psi_{R}^{\dagger}i\sigma^{\nu}D^{\mu}\psi_{R}:\right\rangle ,\label{eq:54h}
\end{equation}
where $\sigma^{\mu}=(1,\boldsymbol{\sigma})$, $D^{\mu}=(\partial_{t},-\partial_{x},-\partial_{y}+ieBx,-\partial_{z})$,
the angular brackets means ensemble average, and the double dots enclosing
the field operators means normal ordering as adopted in \citep{Vasak:1987um,Dong:2020zci}.
In Appendix \ref{sec:uno}, we have compared the results of normal
ordering and un-normal ordering descriptions. For un-normal ordering
description, one should add the vacuum term to the grand partition
function in Eq. (\ref{eq:5.1d}). The un-normal ordering description
is adopted in \citep{Yang:2020mtz}, where the contribution of vacuum
term is considered, resulting in a regular form of $eB$, i.e. $(eB)^{2}\ln(\Lambda^{2}/T^{2}$),
with $\Lambda$ the renormalization scale, and the logarithmic term
of $eB$ disappears. For the QED case, the vacuum term contributes
a similar logarithmic term of the renormalization scale $\Lambda$,
which is related to the beta function of the theory \citep{Dunne:2004nc}.

In the following, we will calculate $T^{00},T^{11},T^{03}$ through
ensemble average with normal ordering description. In order to calculate
Eq. (\ref{eq:54h}), the field operator $\psi_{R}(\boldsymbol{x})$
must be expanded by the orthonormal and complete eigenfunctions in
Eq. (\ref{eq:5.1e}) as follows \citep{Dong:2020zci},
\begin{eqnarray}
\psi_{R}(\boldsymbol{x}) & = & \sum_{k_{y},k_{z}}\left[a_{0}(k_{y},k_{z})\theta(k_{z})\psi_{R0}(k_{y},k_{z};\boldsymbol{x})+b_{0}^{\dagger}(k_{y},k_{z})\theta(-k_{z})\psi_{R0}(k_{y},k_{z};\boldsymbol{x})\right]\nonumber \\
 &  & +\sum_{n,k_{y},k_{z}}\left[a_{n}(k_{y},k_{z})\psi_{Rn+}(k_{y},k_{z};\boldsymbol{x})+b_{n}^{\dagger}(k_{y},k_{z})\psi_{Rn-}(k_{y},k_{z};\boldsymbol{x})\right],\label{eq:54i}
\end{eqnarray}
where $a_{n},a_{n}^{\dagger},b_{n},b_{n}^{\dagger}$ are annihilation
and creation operators for fermions and antifermions. As calculated
in \citep{Dong:2020zci} , the ensemble average of normal ordering
for $a_{n}^{\dagger}a_{n}$ and $b_{n}b_{n}^{\dagger}$ is
\begin{eqnarray}
\left\langle :\theta(k_{z})a_{0}^{\dagger}(k_{y},k_{z})a_{0}(k_{y},k_{z}):\right\rangle  & = & \frac{\theta(k_{z})}{e^{\beta(k_{z}-\mu_{R})}+1},\label{eq:57a}\\
\left\langle :\theta(-k_{z})b_{0}(k_{y},k_{z})b_{0}^{\dagger}(k_{y},k_{z}):\right\rangle  & = & -\frac{\theta(-k_{z})}{e^{\beta(-k_{z}+\mu_{R})}+1},\label{eq:54j}
\end{eqnarray}
\begin{eqnarray}
\left\langle :a_{n}^{\dagger}(k_{y},k_{z})a_{n}(k_{y},k_{z}):\right\rangle  & = & \frac{1}{e^{\beta[E_{n}(k_{z})-\mu_{R}]}+1},\label{eq:57b}\\
\left\langle :b_{n}(k_{y},k_{z})b_{n}^{\dagger}(k_{y},k_{z}):\right\rangle  & = & -\frac{1}{e^{\beta[E_{n}(k_{z})+\mu_{R}]}+1}.\label{eq:54k}
\end{eqnarray}

Substituting Eq. (\ref{eq:54i}) into Eq. (\ref{eq:54h}) and taking
use of Eqs. (\ref{eq:57a})-(\ref{eq:54k}) gives 
\begin{equation}
T^{03}=\frac{eB}{(2\pi)^{2}}\int_{0}^{\infty}dk_{z}k_{z}\bigg(\frac{1}{e^{\beta(k_{z}-\mu_{R})}+1}+\frac{1}{e^{\beta(k_{z}-\mu_{R})}+1}\bigg),\label{eq:510r}
\end{equation}
\begin{equation}
T^{11}=\frac{(eB)^{2}}{2\pi^{2}}\int_{0}^{\infty}dk_{z}\sum_{n=1}^{\infty}\frac{n}{E_{n}}\bigg(\frac{1}{e^{\beta[E_{n}(k_{z})-\mu]}+1}+\frac{1}{e^{\beta[E_{n}(k_{z})+\mu]}+1}\bigg).\label{eq:510s}
\end{equation}
We can see that only the lowest Landau level $(n=0)$ contributes
to $T^{03}$, while $T^{11}$ comes from higher Landau levels $(n>0)$.
Further calculation gives
\begin{equation}
T^{03}=\frac{b}{8\pi^{2}\beta^{4}}\bigg(\frac{\pi^{2}}{6}+\frac{a^{2}}{2}\bigg),\label{eq:510b}
\end{equation}
\begin{equation}
T^{11}=\frac{1}{\beta^{4}}\bigg(1-b\frac{\partial}{\partial b}\bigg)g(a,b).\label{eq:510a}
\end{equation}
We can see that $T^{03}$ is only linear in $b$. In fact, the space
component of the particle number current is $ab/(8\pi^{2}\beta^{3})$,
also linear in $b$ as calculated in \citep{Dong:2020zci}, which
is called chiral magnetic effect. When $a=0$, i.e. the chemical potential
is zero, the particle number current disappears, but a nonzero term
$eBT^{2}/24$ remains in $T^{03}$, which is reasonable since the
same amount of fermions and antifermions move in the same direction
along $z$-axis as discussed in \citep{Dong:2020zci}, resulting in
a zero value of the net particle number current and a nonzero value
of energy current.

For left-handed fermion system, $T^{11}$ can be obtained by simply
replacing $a=\beta\mu_{R}$ with $\beta\mu_{L}$ in Eq. (\ref{eq:510a}).
However, $T^{03}$ for left-handed fermion system can be obtained
by space inversion in Eq. (\ref{eq:510b}) and the result is 
\begin{equation}
-\frac{b}{8\pi^{2}\beta^{4}}\bigg(\frac{\pi^{2}}{6}+\frac{\beta^{2}\mu_{L}^{2}}{2}\bigg).\label{eq:510t}
\end{equation}
The total $T^{03}$ of the chiral fermion system is the summation
of the results of left-handed and right-handed fermion systems, and
the result is $eB\mu\mu_{5}/(2\pi^{2})$ with $\mu=\frac{1}{2}(\mu_{R}+\mu_{L})$
and $\mu_{5}=\frac{1}{2}(\mu_{R}-\mu_{L})$, which is consistent with
\citep{Gao:2012ix}.

From the series expansion of $g(a,b)$, we can obtain 
\begin{eqnarray}
T^{11} & = & \frac{1}{4\pi^{2}\beta^{4}}\bigg[\bigg(\frac{7\pi^{4}}{90}+\frac{\pi^{2}a^{2}}{3}+\frac{a^{4}}{6}\bigg)+\frac{b^{2}\ln b^{2}}{96}+\frac{b^{2}}{24}\ln\bigg(\frac{e^{3/2+C_{1}(a)}}{2G^{6}}\bigg)\nonumber \\
 &  & +\sum_{n=1}^{\infty}\frac{(4n+1)!!}{(4n+4)!!}(4n+2)\mathscr{B}_{2n+2}C_{2n+1}(a)b^{2n+2}\bigg].\label{eq:510u}
\end{eqnarray}
Since $g_{\mu\nu}T^{\mu\nu}=0$ and $T^{11}=T^{22}$, the result for
$T^{33}$ is
\begin{equation}
T^{33}=T^{00}-2T^{11}=\frac{g(a,b)}{\beta^{4}}=p,\label{eq:d1-1}
\end{equation}
so we conclude that $T^{33}$ is just the pressure $p$ of the system. 

\section{Summary}

\label{sec:Summary}In this article, we have studied the thermodynamics
of the chiral fermion system in a uniform magnetic field, where we
ignored the interaction among all fermions. Since the equations of
motion for left-handed fermions and right-handed fermions decouple,
we did all the calculations for the case of right-handed fermion,
which can be generalised to the left-handed case directly. From the
Landau levels of a single right-handed fermion in a uniform magnetic
field, we construct the grand partition function of this thermodynamic
system, through which all intensive quantities can be obtained as
a summation over all Landau levels. Taking use of Abel-Plana formula,
the summation over all Landau levels can be transformed into integrations,
which can be more readily dealt with analytically. We expanded these
thermodynamic quantities as series with respect to a dimensionless
variable $b=2eB/T^{2}$. We find that, the series expansions of energy
density, pressure, magnetization intensity, magnetic susceptibility
contain a singular term with $\ln b^{2}$, which indicates that these
thermodynamic quantities are not analytic at $b=0$. Meanwhile, the
series expansions of particle number density, entropy density and
heat capacity are power series of $b^{2}$, which indicates the analyticity
of them at $b=0$. We plot the curves of these thermodynamic quantities
with respect to $b$ with zero and finite chemical potentials respectively,
and discuss the asymptotic behaviors of these quantities in strong/weak
magnetic field limits and low/high temperature limits. All elements
of energy-momentum tensor are also calculated. We conclude that the
magnetic field can have important influence on the thermodynamics
of chiral fermion system, which may be helpful to study the properties
of the quark gluon plasma state created in high energy heavy ion collisions. 

\section{\textit{Acknowledgments}}

We are grateful to Michael Peskin and Hai-Cang Ren for helpful discussion.
This work was supported by the National Natural Science Foundation
of China under grant Nos. 11890713, 11735007, 11890711.

\appendix

\section{Abel-Plana formula}

\label{sec:Abel-Plana}In mathematics, the Abel-Plana formula is a
summation formula which is discovered by Niels Henrik Abel (1823)
and Giovanni Antonio Amedeo Plana (1820) independently \citep{Butzer:2011}.
It states that: If a function $\mathcal{F}(z)$ is analytic at $\mathrm{Re}\,z\geqslant0$
and $\mathcal{F}(z)\rightarrow0$ as $|z|\rightarrow\infty$ along
positive real axis, then
\begin{equation}
\frac{1}{2}\mathcal{F}(0)+\sum_{n=1}^{\infty}\mathcal{F}(n)=\int_{0}^{\infty}dt\mathcal{F}(t)+i\int_{0}^{\infty}dt\frac{\mathcal{F}(it)-\mathcal{F}(-it)}{e^{2\pi t}-1}.\label{eq:59b}
\end{equation}

In Sec. \ref{sec:Expansion}, we meet with following function,
\begin{equation}
g(a,b)=\frac{b}{4\pi^{2}}\int_{0}^{\infty}ds\left[\frac{1}{2}f(a,s)+\sum_{n=1}^{\infty}f(a,\sqrt{nb+s^{2}})\right],\label{eq:59c}
\end{equation}
where $b>0$ and $f(a,x)$ is defined as
\begin{equation}
f(a,x)=\ln(1+e^{a-x})+\ln(1+e^{-a-x}).\label{eq:a2-1}
\end{equation}
We can see that the function $\mathcal{F}(z)=f(a,\sqrt{zb+s^{2}})$
satisfies $\mathcal{F}(z)\rightarrow0$ as $|z|\rightarrow\infty$
along positive real axis. In order to apply the Abel-Plana formula
to the summation in the integrand in Eq. (\ref{eq:59c}), the function
$\mathcal{F}(z)=f(a,\sqrt{zb+s^{2}})$ must be analytic at $\mathrm{Re}\,z\geqslant0$.
The possible poles appear in $\mathcal{F}(z)$ when 
\begin{equation}
1+e^{\pm a-\sqrt{zb+s^{2}}}=0,\label{eq:510v}
\end{equation}
which implies 
\begin{equation}
z=\frac{1}{b}\left[a^{2}-(2n+1)^{2}\pi^{2}-s^{2}\mp i(4n+2)\pi a\right],\ (n=0,\pm1,\pm2,\cdots).\label{eq:59d}
\end{equation}
In order to make $\mathcal{F}(z)$ analytic at $\mathrm{Re}\,z\geqslant0$,
the poles in Eq. (\ref{eq:59d}) must lie in the range $\mathrm{Re}\,z<0$,
i.e. the inequation $a^{2}-(2n+1)^{2}\pi^{2}-s^{2}<0$ must be satisfied
for any $n$ and $s$, which indicates $-\pi<a<\pi$.

\section{Expansion of $g(a,b)$ at $b=0$}

\label{sec:Expansion-app}Define an auxiliary function $F(a,x)$ as
\begin{equation}
F(a,x)=\int_{0}^{\infty}dsf(a,\sqrt{x^{2}+s^{2}}),\label{eq:ap1}
\end{equation}
where $f(a,x)=\ln(1+e^{a-x})+\ln(1+e^{-a-x})$ as defined in Eq. (\ref{eq:a2}).
Then $g(a,b)$ in Eq. (\ref{eq:5.1l-1}) becomes
\begin{equation}
g(a,b)=\bigg(\frac{7\pi^{2}}{360}+\frac{a^{2}}{12}+\frac{a^{4}}{24\pi^{2}}\bigg)+\frac{b}{4\pi^{2}}\times i\int_{0}^{\infty}dt\frac{F(a,\sqrt{itb})-F(a,\sqrt{-itb})}{e^{2\pi t}-1}.\label{eq:a1}
\end{equation}
Now we will expand $F(a,x)$ at $x=0$ in the following. By the variable
transformation $y=\sqrt{x^{2}+s^{2}}$, $F(a,x)$ can be written as
\begin{equation}
F(a,x)=\int_{|x|}^{\infty}dy\frac{y}{\sqrt{y^{2}-x^{2}}}f(a,y).\label{eq:ap3}
\end{equation}
The factor $y/\sqrt{y^{2}-x^{2}}$ in the integrand in $F(a,x)$ can
be replaced by following Taylor expansion,
\begin{equation}
\frac{y}{\sqrt{y^{2}-x^{2}}}=\sum_{n=0}^{\infty}\frac{(2n-1)!!}{(2n)!!}\frac{x^{2n}}{y^{2n}},\label{eq:ap4}
\end{equation}
where we have defined $(-1)!!=0!!=1$. Then $F(a,x)$ becomes
\begin{equation}
F(a,x)\equiv\sum_{n=0}^{\infty}\frac{(2n-1)!!}{(2n)!!}x^{2n}d_{n}(a,x),\label{eq:z1}
\end{equation}
where we have defined $d_{n}(a,x)$ as

\begin{equation}
d_{n}(a,x)=\int_{|x|}^{\infty}dy\frac{1}{y^{2n}}f(a,y).\label{eq:510w}
\end{equation}

The derivative of $d_{n}(a,x)$ with respect to $x$ is
\begin{equation}
d_{n}^{\prime}(a,x)=-\frac{|x|}{x^{2n+1}}[f(a,x)+x]+\frac{1}{x^{2n-1}}.\label{eq:ap5}
\end{equation}
Note that the factor $f(a,x)+x$ in Eq. (\ref{eq:ap5}) is an even
function of $x$ and can be expanded at $x=0$ as follows
\begin{equation}
f(a,x)+x=\sum_{k=0}^{\infty}c_{k}(a)x^{2k},\label{eq:z4}
\end{equation}
From $d_{n}^{\prime}(a,x)$ we can derive $d_{n}(a,x)$ as
\begin{equation}
d_{n}(a,x)=-|x|\sum_{k=0}^{\infty}\frac{c_{k}(a)}{2k-2n+1}x^{2k-2n}+\left\{ \begin{array}{cc}
C_{1}(a)+\frac{1}{2}\ln x^{2}, & n=1\\
C_{n}(a)-\frac{1}{2(n-1)}x^{2-2n}, & n\neq1
\end{array}\right.,\label{eq:510x}
\end{equation}
where $C_{n}(a)$ are independent of $x$ which will be calculated
later.

Taking use of following two identities,
\begin{equation}
\sum_{n=2}^{\infty}\frac{(2n-1)!!}{(2n)!!}\frac{1}{2(n-1)}=\frac{1}{2}\ln2+\frac{1}{4},\label{eq:510y}
\end{equation}
\begin{equation}
\sum_{n=0}^{\infty}\frac{(2n-1)!!}{(2n)!!}\frac{1}{2n-2k-1}=0,\ \ (k=0,1,2,\cdots),\label{eq:510z}
\end{equation}
we can obtain $F(a,x)$ in Eq. (\ref{eq:z1}) as
\begin{equation}
F(a,x)=C_{0}(a)+\frac{1}{4}x^{2}\ln x^{2}+\bigg(\frac{1}{4}-\frac{1}{2}\ln2+\frac{1}{2}C_{1}(a)\bigg)x^{2}+\sum_{n=2}^{\infty}\frac{(2n-1)!!}{(2n)!!}C_{n}(a)x^{2n}.\label{eq:510a1}
\end{equation}

In the following we will extract $C_{n}(a)$ in $d_{n}(a,x)$. When
$n=0$, we have $C_{0}(a)=d_{0}(a,0)=\pi^{2}/6+a^{2}/2$. When $n>0$,
through integration by parts we have
\begin{eqnarray}
d_{n}(a,x) & = & \sum_{k=0}^{2n-2}\frac{(2n-k-2)!}{(2n-1)!}\frac{1}{x^{2n-k-1}}\frac{d^{k}}{dx^{k}}f(a,x)-\frac{\ln x}{(2n-1)!}\frac{d^{2n-1}}{dx^{2n-1}}f(a,x)\nonumber \\
 &  & -\frac{1}{(2n-1)!}\int_{x}^{\infty}dy\ln y\frac{d^{2n}}{dy^{2n}}f(a,y).\label{eq:510a2}
\end{eqnarray}
Taking use of Eq. (\ref{eq:z4}), we obtain
\begin{equation}
C_{n}(a)=-\delta_{n,1}-\frac{1}{(2n-1)!}\int_{0}^{\infty}dy\ln y\frac{d^{2n}}{dy^{2n}}f(a,y).\label{eq:58a}
\end{equation}

From Eq. (\ref{eq:a1}), we can see that the terms with $x^{4n}\ (n\geqslant0)$
in $F(a,x)$ do not contribute to $g(a,b)$, so we can express $F(a,x)$
as follows,
\begin{eqnarray}
F(a,x) & = & \frac{1}{4}x^{2}\ln x^{2}+\bigg(\frac{1}{4}-\frac{1}{2}\ln2+\frac{1}{2}C_{1}(a)\bigg)x^{2}\nonumber \\
 &  & +\sum_{n=1}^{\infty}\frac{(4n+1)!!}{(4n+2)!!}C_{2n+1}(a)x^{4n+2}+\sum_{n=0}^{\infty}(\mathrm{terms\ with\ }x^{4n}).\label{eq:z3}
\end{eqnarray}
Substituting Eq. (\ref{eq:z3}) into Eq. (\ref{eq:a1}), we have
\begin{eqnarray}
g(a,b) & = & \bigg(\frac{7\pi^{2}}{360}+\frac{a^{2}}{12}+\frac{a^{4}}{24\pi^{2}}\bigg)-\frac{b^{2}\ln b^{2}}{384\pi^{2}}-\frac{b^{2}}{96\pi^{2}}\ln\bigg(\frac{e^{1+C_{1}(a)}}{2G^{6}}\bigg)\nonumber \\
 &  & -\frac{1}{2\pi^{2}}\sum_{n=1}^{\infty}\frac{(4n+1)!!}{(4n+4)!!}\mathscr{B}_{2n+2}C_{2n+1}(a)b^{2n+2},\label{eq:q1}
\end{eqnarray}
where we have used following integrations,
\begin{equation}
\int_{0}^{\infty}dt\frac{t\ln t}{e^{2\pi t}-1}=\frac{1}{24}-\frac{1}{2}\ln G,\label{eq:510a3}
\end{equation}
\begin{equation}
\int_{0}^{\infty}dt\frac{t^{2n+1}}{e^{2\pi t}-1}=(-1)^{n}\frac{\mathscr{B}_{2n+2}}{4n+4},\ \ (n\geqslant0),\label{eq:510a4}
\end{equation}
with Glaisher number $G=1.282427...$ and Bernoulli numbers $\mathscr{B}_{n}$
defined as
\begin{equation}
\frac{t}{e^{t}-1}=\sum_{n=0}^{\infty}\frac{\mathscr{B}_{n}}{n!}t^{n}.\label{eq:510a5}
\end{equation}
We list some Bernoulli number $\mathscr{B}_{2n+2}$ as follows,
\begin{equation}
\mathscr{B}_{2}=\frac{1}{6},\ \ \mathscr{B}_{4}=-\frac{1}{30},\ \ \mathscr{B}_{6}=\frac{1}{42},\ \ \mathscr{B}_{8}=-\frac{1}{30}.\label{eq:510a6}
\end{equation}

\section{Asymptotic behavior of $C_{2n+1}(a)$ as $a\rightarrow\infty$}

\label{sec:Aymp-c(2n+1)}To study the aymptotic behavior of $C_{2n+1}(a)$
as $a\rightarrow\infty$, we rewrite $C_{2n+1}(a)$ as
\begin{equation}
C_{2n+1}(a)=-\delta_{n,0}+\frac{1}{(4n+1)!}\frac{d^{4n+1}}{da^{4n+1}}\int_{0}^{\infty}dy\ln y\left(\frac{1}{e^{y+a}+1}-\frac{1}{e^{y-a}+1}\right).\label{eq:510a7}
\end{equation}
Taking use of following integration formula,
\begin{equation}
\int_{0}^{\infty}dy\ln y\left(\frac{1}{e^{y+a}+1}-\frac{1}{e^{y-a}+1}\right)=\gamma a+\left\{ \frac{d}{ds}\left[\mathrm{Li}_{s}(-e^{a})-\mathrm{Li}_{s}(-e^{-a})\right]\right\} _{s=1},\label{eq:510a8}
\end{equation}
we have
\begin{equation}
C_{2n+1}(a)=(\gamma-1)\delta_{n,0}+\frac{1}{(4n+1)!}\frac{d^{4n+1}}{da^{4n+1}}\left\{ \frac{d}{ds}\left[\mathrm{Li}_{s}(-e^{a})-\mathrm{Li}_{s}(-e^{-a})\right]\right\} _{s=1},\label{eq:59a-1}
\end{equation}
where $\mathrm{Li}_{s}(z)$ is the polylogarithm function defined
as
\begin{equation}
\mathrm{Li}_{s}(z)=\sum_{k=1}^{\infty}\frac{z^{k}}{k^{s}}.\label{eq:510a9}
\end{equation}
This definition of $\mathrm{Li}_{s}(z)$ is valid for arbitrary complex
number $s$ and for $|z|<1$, and it can be extended to $|z|\geqslant1$
by the process of analytic continuation. Taking use of following asymptotic
formulas for $\mathrm{Li}_{s}(z)$,

\begin{equation}
\lim_{z\rightarrow0}\mathrm{Li}_{s}(z)=z,\label{eq:510b1}
\end{equation}
\begin{equation}
\lim_{a\rightarrow\infty}\mathrm{Li}_{s}(-e^{a})=-\frac{a^{s}}{\Gamma(s+1)},\ \ (s\neq-1,-2,-3,\cdots),\label{eq:510b2}
\end{equation}
we have
\begin{equation}
\lim_{a\rightarrow\infty}\left[\frac{d}{ds}\left(\mathrm{Li}_{s}(-e^{a})-\mathrm{Li}_{s}(-e^{-a})\right)\right]_{s=1}=(1-\gamma)a-a\ln a,\label{eq:510b3}
\end{equation}
where we have used $\Gamma^{\prime}(2)=1-\gamma$. So we obtain
\begin{eqnarray}
\lim_{a\rightarrow\infty}C_{2n+1}(a) & = & (\gamma-1)\delta_{n,0}+\frac{1}{(4n+1)!}\frac{d^{4n+1}}{da^{4n+1}}\left[(1-\gamma)a-a\ln a\right]\nonumber \\
 & = & -\frac{1}{(4n+1)!}\frac{d^{4n+1}}{da^{4n+1}}\left(a\ln a\right)\nonumber \\
 & = & \left\{ \begin{array}{cc}
-1-\ln a, & n=0\\
\frac{1}{4n(4n+1)}a^{-4n}, & n>0
\end{array}\right.,\label{eq:510b4}
\end{eqnarray}
which shows that $C_{1}(a)\rightarrow-\infty$ and $C_{2n+1}(a)\rightarrow0$
$(n>0)$ as $a\rightarrow\infty$. Furthermore, we have
\begin{equation}
\lim_{a\rightarrow\infty}C_{2n+1}^{\prime}(a)=-\frac{1}{4n+1}a^{-4n-1}\rightarrow0,\ \ (n\geqslant0).\label{eq:510b5}
\end{equation}

\section{Expansion of $C_{2n+1}(a)$ at $a=0$}

\label{sec:C(2n+1)at-0}To expand $C_{2n+1}(a)$ at $a=0$, we use
following expression of $C_{2n+1}(a)$,
\begin{equation}
C_{2n+1}(a)=-\delta_{n,0}+\frac{1}{(4n+1)!}\int_{0}^{\infty}dy\ln y\frac{d^{4n+1}}{dy^{4n+1}}\left(\frac{1}{e^{y+a}+1}+\frac{1}{e^{y-a}+1}\right).\label{eq:512c}
\end{equation}

When $n>0$, we have

\begin{align}
C_{2n+1}(a)= & \frac{2}{(4n+1)!}\int_{0}^{\infty}dy\ln y\frac{d^{4n+1}}{dy^{4n+1}}\sum_{k=0}^{\infty}\frac{a^{2k}}{(2k)!}\frac{d^{2k}}{dy^{2k}}\left(\frac{1}{e^{y}+1}\right)\nonumber \\
= & -\frac{2}{(4n+1)!}\sum_{k=0}^{\infty}\frac{a^{2k}}{(2k)!}\int_{0}^{\infty}dy\frac{1}{y}\frac{d^{4n+2k}}{dy^{4n+2k}}\left(\frac{1}{e^{y}+1}\right).\label{eq:512a}
\end{align}
In order to calculate the integration in the second line of Eq. (\ref{eq:512a}),
we take use of following identities,
\begin{equation}
\Gamma(s)\mathrm{Li}_{s}(-e^{x})=-\int_{0}^{\infty}dy\frac{1}{y^{1-s}}\frac{1}{e^{y-x}+1},\ \ (\mathrm{Re\,}s>0),\label{eq:512p}
\end{equation}
\begin{equation}
\frac{d}{dx}\mathrm{Li}_{s}(-e^{x})=\mathrm{Li}_{s-1}(-e^{x}),\label{eq:512q}
\end{equation}
\begin{equation}
\lim_{s\rightarrow0}\Gamma(s)\mathrm{Li}_{s-2n}(-1)=\left(2^{2n+1}-1\right)\zeta^{\prime}(-2n),\ \ (n=1,2,3,\cdots),\label{eq:512r}
\end{equation}
then we can obtain
\begin{eqnarray}
 &  & \int_{0}^{\infty}dy\frac{1}{y}\frac{d^{4n+2k}}{dy^{4n+2k}}\left(\frac{1}{e^{y}+1}\right)\nonumber \\
 & = & \lim_{s\rightarrow0}\int_{0}^{\infty}dy\frac{1}{y^{1-s}}\left(\frac{d^{4n+2k}}{dx^{4n+2k}}\frac{1}{e^{y-x}+1}\right)_{x=0}\nonumber \\
 & = & -\lim_{s\rightarrow0}\Gamma(s)\left[\frac{d^{4n+2k}}{dx^{4n+2k}}\mathrm{Li}_{s}(-e^{x})\right]_{x=0}\nonumber \\
 & = & -\lim_{s\rightarrow0}\Gamma(s)\mathrm{Li}_{s-4n-2k}(-1)\nonumber \\
 & = & \left(1-2^{4n+2k+1}\right)\zeta^{\prime}(-4n-2k).\label{eq:512s}
\end{eqnarray}
So Eq. (\ref{eq:512a}) becomes
\begin{equation}
C_{2n+1}(a)=\frac{2}{(4n+1)!}\sum_{k=0}^{\infty}\left(2^{4n+2k+1}-1\right)\zeta^{\prime}(-4n-2k)\frac{a^{2k}}{(2k)!}.\label{eq:512t}
\end{equation}
When $n=0$ in Eq. (\ref{eq:512c}), similar calculation gives
\begin{equation}
C_{1}(a)=\ln4+\gamma-1+2\sum_{k=0}^{\infty}\left(2^{2k+1}-1\right)\zeta^{\prime}(-2k)\frac{a^{2k}}{(2k)!}.\label{eq:512u}
\end{equation}

The general expression of $C_{2n+1}(a)$ for $n\geqslant0$ is
\begin{equation}
C_{2n+1}(a)=(\ln4+\gamma-1)\delta_{n,0}+\frac{2}{(4n+1)!}\sum_{k=0}^{\infty}\left(2^{4n+2k+1}-1\right)\zeta^{\prime}(-4n-2k)\frac{a^{2k}}{(2k)!}.\label{eq:512o}
\end{equation}
Especially, when $a=0$, we have
\begin{equation}
C_{2n+1}(0)=(\ln4+\gamma-1)\delta_{n,0}+\frac{2\zeta^{\prime}(-4n)}{(4n+1)!}\left(2^{4n+1}-1\right).\label{eq:512v}
\end{equation}

\section{Asymptotic behavior of $g(a,b)$ as $b\rightarrow\infty$}

\label{sec:asymptotic}To study the asymptotic behavior of $g(a,b)$
as $b\rightarrow\infty$, we may rewrite $g(a,b)$ in Eq. (\ref{eq:5.1l-1})
as
\begin{equation}
g(a,b)=\bigg(\frac{7\pi^{2}}{360}+\frac{a^{2}}{12}+\frac{a^{4}}{24\pi^{2}}\bigg)+\frac{1}{4\pi^{2}}\times i\int_{0}^{\infty}ds\int_{0}^{\infty}dt\frac{f(a,\sqrt{it+s^{2}})-f(a,\sqrt{-it+s^{2}})}{e^{2\pi t/b}-1}.\label{eq:q4}
\end{equation}
where $f(a,x)=\ln(1+e^{a-x})+\ln(1+e^{-a-x})$ as defined in Eq. (\ref{eq:a2}).
As $z\rightarrow0$, we can expand $1/(e^{z}-1)$ as follows,

\begin{equation}
\frac{1}{e^{z}-1}=\frac{1}{z}-\frac{1}{2}+\frac{z}{12}-\frac{z^{3}}{720}+\frac{z^{5}}{30240}+\cdots\label{eq:f1}
\end{equation}
So as $b\rightarrow\infty$ (i.e. $1/b\rightarrow0$), $g(a,b)$ becomes
\begin{eqnarray}
g(a,b) & = & \bigg(\frac{7\pi^{2}}{360}+\frac{a^{2}}{12}+\frac{a^{4}}{24\pi^{2}}\bigg)+\frac{b}{8\pi^{3}}\times i\int_{0}^{\infty}dt\frac{1}{t}\int_{0}^{\infty}ds\left[f(a,\sqrt{it+s^{2}})-f(a,\sqrt{-it+s^{2}})\right]\nonumber \\
 &  & -\frac{1}{8\pi^{2}}\times i\int_{0}^{\infty}dt\int_{0}^{\infty}ds\left[f(a,\sqrt{it+s^{2}})-f(a,\sqrt{-it+s^{2}})\right]+\mathcal{O}(\frac{1}{b}).\label{eq:510c1}
\end{eqnarray}
Fortunately, the two integrations in Eq. (\ref{eq:510c1}) can be
analytically integrated out as follows,
\begin{equation}
i\int_{0}^{\infty}dt\frac{1}{t}\int_{0}^{\infty}ds\left[f(a,\sqrt{it+s^{2}})-f(a,\sqrt{-it+s^{2}})\right]=\frac{\pi^{3}}{6}+\frac{\pi}{2}a^{2},\label{eq:513a}
\end{equation}
\begin{equation}
i\int_{0}^{\infty}dt\int_{0}^{\infty}ds\left[f(a,\sqrt{it+s^{2}})-f(a,\sqrt{-it+s^{2}})\right]=\frac{7\pi^{4}}{45}+\frac{2\pi^{2}}{3}a^{2}+\frac{1}{3}a^{4},\label{eq:513h}
\end{equation}
So we have
\begin{equation}
\lim_{b\rightarrow\infty}g(a,b)=\frac{b}{48\pi^{2}}\left(\pi^{2}+3a^{2}\right).\label{eq:513b}
\end{equation}

\section{Normal ordering and un-normal ordering}

\label{sec:uno}The explicit form of energy density $\varepsilon$
expressed by a summation of all quantum states in Eq. (\ref{eq:54a})
can also be derived from the ensemble average of Normal Ordering (NO)
of the energy density operator as described in Sec. VI. 

If we adopt Un-Normal Ordering (UNO), then Eqs. (\ref{eq:54j}) and
(\ref{eq:54k}) become
\begin{eqnarray}
\left\langle \theta(-k_{z})b_{0}(k_{y},k_{z})b_{0}^{\dagger}(k_{y},k_{z})\right\rangle  & = & \theta(-k_{z})\left[1-\frac{1}{e^{\beta(-k_{z}+\mu_{R})}+1}\right],\label{eq:57c}\\
\left\langle b_{n}(k_{y},k_{z})b_{n}^{\dagger}(k_{y},k_{z})\right\rangle  & = & 1-\frac{1}{e^{\beta[E_{n}(k_{z})+\mu_{R}]}+1}.\label{eq:57d}
\end{eqnarray}
This UNO description is also used in recent articles \citep{Sheng:2017lfu,Gao:2019zhk}.
Now the ensemble average of the energy density operator becomes
\begin{eqnarray}
\varepsilon_{\mathrm{un}} & = & \frac{1}{V}\sum_{k_{y},k_{z}}\left[k_{z}\theta(k_{z})\frac{1}{e^{\beta(k_{z}-\mu_{R})}+1}+(-k_{z})\theta(-k_{z})\left(\frac{1}{e^{\beta(-k_{z}+\mu_{R})}+1}-1\right)\right]\nonumber \\
 &  & +\frac{1}{V}\sum_{n=1}^{\infty}\sum_{k_{y},k_{z}}E_{n}(k_{z})\left[\frac{1}{e^{\beta[E(n,k_{z})-\mu_{R}]}+1}+\frac{1}{e^{\beta[E(n,k_{z})+\mu_{R}]}+1}-1\right],\label{eq:57e}
\end{eqnarray}
where the subscript ``un'' of $\varepsilon_{\mathrm{un}}$ means
un-normal ordering. We can see that UNO description leads to the infinite
vacuum energy in $\varepsilon_{\mathrm{un}}$. Comparing Eq. (\ref{eq:54a})
and Eq. (\ref{eq:57e}), it is equivalent to do following regularization,
\begin{equation}
\varepsilon_{\mathrm{un}}=\varepsilon-\lim_{\mu_{R}\rightarrow\infty}\varepsilon.\label{eq:57f}
\end{equation}
Taking use of two dimensionless variables $a=\beta\mu_{R}$ and $b=2eB\beta^{2}$
instead of $\mu_{R},\beta,eB$, Eq. (\ref{eq:57f}) can be rewritten
as
\begin{equation}
\varepsilon_{\mathrm{un}}(a,b)=\varepsilon(a,b)-\lim_{a\rightarrow\infty}\varepsilon(a,b).\label{eq:57g}
\end{equation}
From the expression of $\varepsilon(a,b)$ in Eq. (\ref{eq:a11}),
Eq. (\ref{eq:57g}) becomes
\begin{equation}
\varepsilon_{\mathrm{un}}(a,b)=\varepsilon_{0}+\frac{1}{2\pi^{2}\beta^{4}}\bigg[\bigg(\frac{\pi^{2}a^{2}}{2}+\frac{a^{4}}{4}\bigg)+\sum_{n=0}^{\infty}\frac{(4n+1)!!}{(4n+4)!!}(4n+1)\mathscr{B}_{2n+2}C_{2n+1}^{\mathrm{un}}(a)b^{2n+2}\bigg],\label{eq:57h}
\end{equation}
where $C_{2n+1}^{\mathrm{un}}(a)$ is defined as
\begin{equation}
C_{2n+1}^{\mathrm{un}}(a)=C_{2n+1}(a)-\lim_{a\rightarrow\infty}C_{2n+1}(a).\label{eq:57i}
\end{equation}
and $\varepsilon_{0}$ is just the familiar infinite vacuum energy,
as in the free field theory \citep{Peskin:1995,Kapusta:2006}. This
infinite vacuum energy can not be detected experimentally, since only
the energy difference from the ground state can be observable. It
is worth pointing out that the term $b^{2}\ln b^{2}$ in $\varepsilon$
disappears in $\varepsilon_{\mathrm{un}}$ and the $b^{2}$ order
of $\varepsilon_{\mathrm{un}}$ is consistent with \citep{Yang:2020mtz}.

In Eq. (\ref{eq:510b4}), we obtained the asymptotic behavior of $C_{2n+1}(a)$
as $a\rightarrow\infty$,
\begin{equation}
\lim_{a\rightarrow\infty}C_{2n+1}(a)=\left\{ \begin{array}{cc}
-1-\ln a, & n=0\\
\frac{1}{4n(4n+1)}a^{-4n}, & n>0
\end{array}\right.,\label{eq:512x}
\end{equation}
which shows that $C_{1}(a)\rightarrow-\infty$ and $C_{2n+1}(a)\rightarrow0$
$(n>0)$ as $a\rightarrow\infty$. So we have $C_{2n+1}^{\mathrm{un}}(a)=C_{2n+1}(a)$
for $n>0$. For $n=0$, another expression of $C_{1}(a)$ is useful
\begin{equation}
C_{1}(a)=\gamma-1-\int_{0}^{\infty}dy\frac{1}{y}\bigg(\frac{1}{e^{y-a}+1}+\frac{1}{e^{y+a}+1}-\frac{1}{e^{y}}\bigg),\label{eq:512y}
\end{equation}
which leads to
\begin{equation}
C_{1}^{\mathrm{un}}(a)=\int_{0}^{\infty}dy\frac{1}{y}\bigg(1-\frac{1}{e^{y-a}+1}-\frac{1}{e^{y+a}+1}\bigg).\label{eq:512z}
\end{equation}
Note that $C_{1}^{\mathrm{un}}(a)$ is a positive infinity which is
expected since $C_{1}(a)\rightarrow-\infty$ as $a\rightarrow\infty$.
This expression of $C_{1}^{\mathrm{un}}(a)$ can also be obtained
in \citep{Yang:2020mtz} where the authors adopt dimensional regularization
to deal with the divergence and obtain a regular term $\ln(\Lambda^{2}/T^{2}$)
instead of the singular term $\ln(eB/T^{2})$ in the energy density
$\varepsilon_{\mathrm{un}}$, with the renormalization scale $\Lambda$.

Since only the derivative of $C_{2n+1}(a)$ appears in the expression
of the particle number density $n$ in Eq. (\ref{eq:54p}), and $C_{2n+1}^{\prime}(a)\rightarrow0$
as $a\rightarrow\infty$ as calculated in Eq. (\ref{eq:510b5}), i.e.
$C_{2n+1}^{\mathrm{un}\prime}(a)=C_{2n+1}^{\prime}(a)$, so the UNO
and NO descriptions can obtain the same expression of the particle
number density $n$, which is also consistent with \citep{Yang:2020mtz}.

\bibliographystyle{apsrev}
\addcontentsline{toc}{section}{\refname}\bibliography{ref-1}

\end{document}